\documentclass[twocolumn]{aastex62}

\received{January 25, 2020}
\revised{...}
\accepted{...} 
\submitjournal{ApJ}

\shorttitle{LPGRE and Shock Wave Properties}
\shortauthors{Kouloumvakos et al. 2020}

\begin{document}

\title{ Evidence for a Coronal Shock Wave Origin for Relativistic Protons Producing Solar Gamma-Rays and Observed by Neutron Monitors at Earth }

\correspondingauthor{Athanasios Kouloumvakos}
\email{akouloumvako@irap.omp.eu}

\author{Athanasios Kouloumvakos}

\affiliation{IRAP, Universit\'e Toulouse III - Paul Sabatier,
CNRS, CNES, Toulouse, France}

\author{Alexis P. Rouillard}
\affiliation{IRAP, Universit\'e Toulouse III - Paul Sabatier,
CNRS, CNES, Toulouse, France}

\author{Gerald H. Share}
\affiliation{Astronomy Department, University of Maryland, College Park, MD 20740, USA}

\author{Illya Plotnikov}
\affiliation{IRAP, Universit\'e Toulouse III - Paul Sabatier,
CNRS, CNES, Toulouse, France}

\author{Ronald Murphy}
\affiliation{Naval Research Laboratory, Washington, DC 20375, USA}

\author{Athanasios Papaioannou}
\affiliation{Institute for Astronomy, Astrophysics, Space Applications and Remote Sensing (IAASARS),\\ National Observatory of Athens, I. Metaxa \& Vas. Pavlou St., GR-15236, Penteli, Greece}

\author{Yihong Wu}
\affiliation{IRAP, Universit\'e Toulouse III - Paul Sabatier,
CNRS, CNES, Toulouse, France}


\begin{abstract}

We study the solar eruptive event on 2017 September 10 that produced long-lasting $>$100 MeV $\gamma$-ray emission and a ground level enhancement (GLE72). The origin of the high-energy ions producing late-phase gamma-ray emission (LPGRE) is still an open question, but a possible explanation is proton acceleration at coronal shocks produced by coronal mass ejections. We examine a common shock acceleration origin for both the LPGRE and GLE72. The $\gamma$-ray emission observed by the Fermi-Large Area Telescope exhibits a weak impulsive phase, consistent with that observed in hard X-and $\gamma$-ray line flare emissions, and what appear to be two distinct stages of LPGRE. From a detailed modeling of the shock wave, we derive the 3D distribution and temporal evolution of the shock parameters, and we examine the shock wave magnetic connection with the visible solar disk. The evolution of shock parameters on field lines returning to the visible disk, mirrors the two stages of LPGRE. We find good agreement between the time history of $>$100 MeV $\gamma$-rays and one produced by a basic shock acceleration model. The time history of shock parameters magnetically mapped to Earth agrees with the rates observed by the Fort Smith neutron monitor during the first hour of the GLE72 if we include a 30\% contribution of flare-accelerated protons during the first 10 minutes, having a release time following the time history of nuclear $\gamma$-rays. Our analysis provides compelling evidence for a common shock origin for protons producing the LPGRE and most of the particles observed in GLE72.

\end{abstract}


\section{Introduction} \label{sec:intro}

Observations of $>$100~MeV $\gamma$-ray emission during major solar eruptions provide a good opportunity to test the different mechanisms that are potentially responsible for the acceleration of protons to $>$300~MeV energies in the solar corona. When such high-energy protons propagate sunward and impact the chromosphere, pions ($\pi$) can be produced. The decay of neutral pions ($\pi^0$) produces $\gamma$-ray emission that dominates the $\gamma$-ray spectrum at energies $>$50 MeV. The Large Area Telescope \citep[LAT:][]{Atwood2009} on-board the Fermi satellite performs observations of $>$100~MeV $\gamma$-ray emission in a limited duty cycle for solar observations. Due to its high sensitivity, the instrument also measures significantly weaker $\gamma$-ray emissions that can last for hours beyond the flare impulsive phase. Such Late Phase Gamma Ray Emission\footnote{see Section 2 for the detailed definition} (LPGRE) which has been observed for over 25 years \citep{Ryan2000,Chupp2009,Ackermann2014,Ajello2014,Share2018}, poses a challenge to the current flare models since the long-lasting emission could imply that particle acceleration continues well after the impulsive phase of solar flares \citep[e.g.][]{Ryan1991,Kahler2018}.

The most important and challenging issue for the interpretation of the long duration $>$100 MeV $\gamma$-ray emission is the relative roles played by the flare and of the shock waves driven by coronal mass ejections (CMEs) in the acceleration of protons beyond 300 MeV. Recent studies examined the role of shock wave acceleration of such protons through statistical analysis incorporating the flare, CME, and other radio-related or solar energetic particle (SEP)-related characteristics of the long duration $\gamma$-ray events measured by  Fermi-LAT \citep{Gopalswamy2018b, Share2018, Winter2018}. The results from those statistical analyses show that LPGRE events are: (i) associated with fast and wide CMEs, (ii) not necessarily associated with the strong soft X-ray flares (but are associated with flares with hard X-ray emission $>$100~keV), and (iii) related to long-duration interplanetary (IP) Type-II radio bursts. Therefore, CME-driven shock waves are strong candidate accelerators for the particles that produce the LPGRE. The average properties of CMEs or Type-II radio bursts can only capture a fraction of the spatial and temporal properties of large-scale shock waves. Therefore, more detailed observations and 3D modeling of the evolution of the shock waves, from their initiation point to their propagation into the IP medium, is required.  

Detection by Fermi-LAT of $>$100 MeV emission requires that ion-nuclear interactions take place on the solar disk visible from Earth. On occasion, high-energy $\gamma$-ray emission has been observed when the solar flare deemed associated with an event originates in an active region (AR) situated on the far side of the Sun \citep{Ackermann2017,Plotnikov2017}. In those events, if the $>$300~MeV protons were accelerated at the flare, they would have to find a way to propagate towards and impact the visible disk. Significantly, it was recently shown that the onset of high-energy $\gamma$-ray emission on 2014 September 01 \citep{Pesce2015,Ackermann2017,Share2018}, from a flare well behind the East limb, occurred when a strong shock wave connected magnetically to the visible disk \citep{Plotnikov2017,Jin2018} and none of the magnetic field lines were found to connect the AR hosting the flare with the visible disk. This supports the role of shock waves as a primary accelerating agent of high-energy particles. Additionally, other comparative studies between the properties of shock waves and SEPs suggest a link between the intensity of the SEP events measured in situ and the strength of the shock waves as quantified by their Mach number \citep[][]{Rouillard2016, Kouloumvakos2019}.

Here, we perform a detailed 3D study of the evolution of the CME shock from the limb flare on 2017 September 10 using the methods described in \cite{Kouloumvakos2019} and estimate how the shock parameters may affect the properties of particles imparted to field lines returning to the visible solar disk with high-enough energies to produce $\pi$-decay $\gamma$-rays in the solar atmosphere. We find a good correlation between our estimates of the shock parameters and the time history of the $\gamma$-rays observed by Fermi/LAT. 

In Section~\ref{sec:ObsAnal} we start the analysis by providing a summary of the $\gamma$-ray, X-ray, and CME observations, setting the foundation for our detailed 3D modeling that we discuss in Section~\ref{sec:ShockModel}. In that section we perform a reconstruction and a 3D modeling of the shock wave and we register its parameters at the field lines that connect the shock surface to the visible disk. In Section~\ref{sec:GrayShock}, we compare the temporal evolution of the shock parameters at regions magnetically connected to the visible solar disk with the measured characteristics of the LPGRE.  We also include effects due to magnetic mirroring that acts to prevent accelerated particles from returning to the solar atmosphere  \citep[][]{Hudson2018}. We show for the first time that there is a consistent relationship between the evolution of the shock and its magnetic connection to the visible disk with the evolution of the $\gamma$-ray emission for the whole extent of the late-phase of $>$100 MeV emission. In Section~\ref{sec:GLEShock} we apply the same methods to estimate the time history of shock accelerated particles observed during ground level enhancement (GLE). We discuss the results of our study in Section~\ref{sec:Discussion}, with emphasis on the what appears to be a common shock origin for both LPGRE events and GLEs and future studies.

\begin{figure}[ht!]
\centering
\includegraphics[scale=0.39]{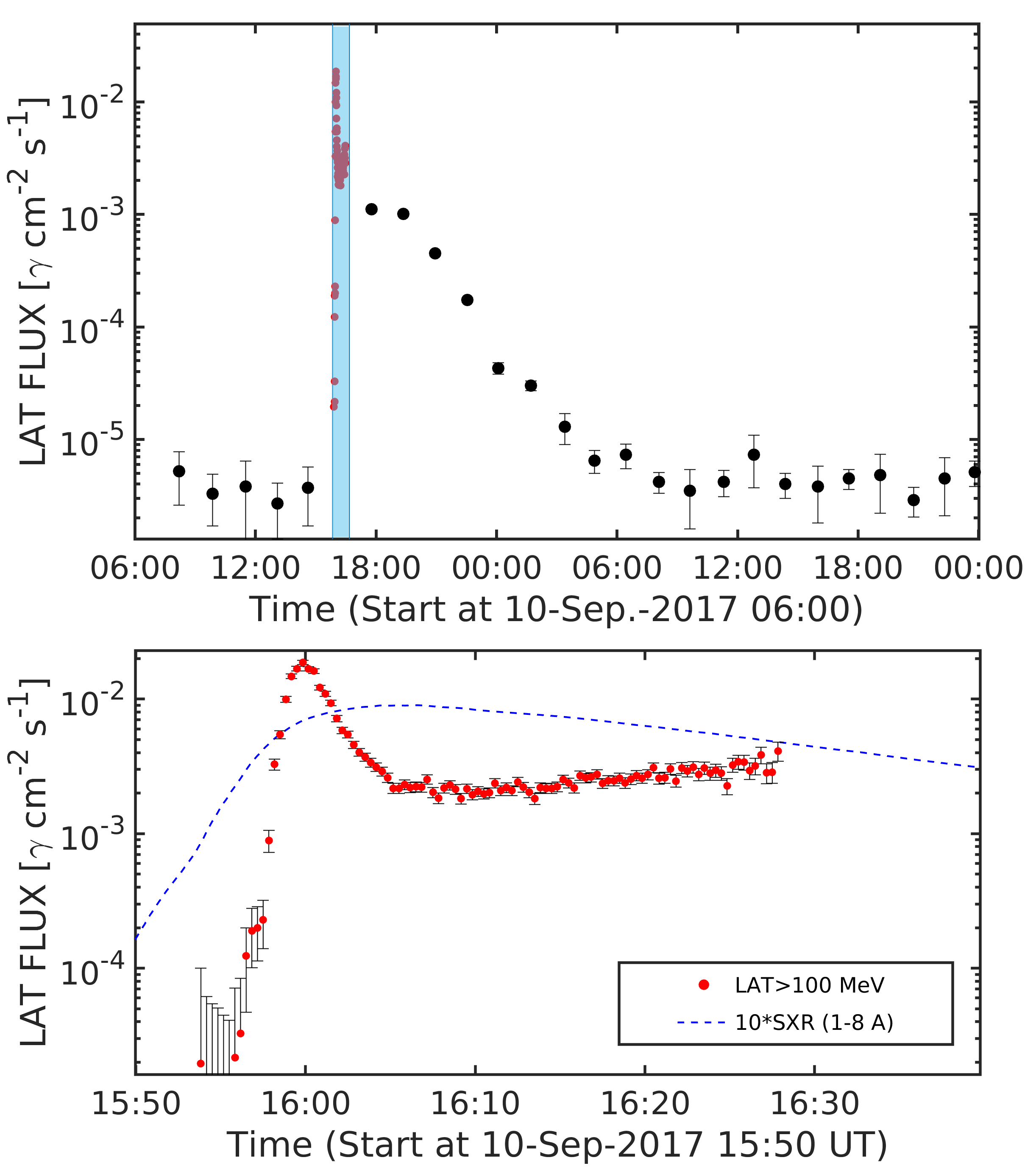}
\caption{ Fermi-LAT observations of the $>$100 MeV solar $\gamma$-ray flux during the 2017 September 10 event. Top: Time history of the $>$100 MeV solar $\gamma$-ray flux for an extended time interval covering the whole event. The blue shaded region depicts the focused time interval shown in the bottom panel. Bottom: Time history of the $>$100 MeV solar $\gamma$-ray flux (red points; measured at a cadence of 20 seconds) and soft X-ray (blue line) emission, from 15:50 to 16:40~UT. \label{fig:FLATLC} }
\end{figure}

\section{Observations}\label{sec:ObsAnal}

\subsection{Flare and CME Observations}

On 2017 September 10 an X8.2 class flare commenced at 15:35~UT near the edge of the western solar limb ($>$W90\footnote{\url{https://cdaw.gsfc.nasa.gov/CME_list/sepe/}}), in NOAA AR12673. This eruptive event was accompanied by a very fast ($\mathrm{>3500~km\,s^{-1}}$) and wide CME \citep[see][]{Veronig2018,Gopalswamy2018a}, as well as the global evolution of an Extreme Ultra-Violet (EUV) wave in the low corona \citep{Liu2018}, and the formation of a fast and wide white-light (WL) shock wave \citep[e.g.][]{Gopalswamy2018a}. The event was also observed at radio wavelengths by the Expanded Owens Valley Solar Array (EOVSA) \citep{Gary2018} and the Low-Frequency Array (LOFAR) \citep{Morosan2019}. High-energy particles were recorded at different points within the heliosphere near L1 and STEREO-A (STA) which was located $\sim$128$^{\circ}$ ahead of the Earth, indicating a widespread SEP event. The protons in this event reached GeV energies that were recorded by neutron monitors on Earth. This was the second GLE of solar cycle 24 and the 72nd \citep[i.e. GLE72; see][]{Mishev2018,Kurt2019} event since records began. Additionally, SEPs were also recorded on the surface of Mars leading to a GLE on another planet \citep[][]{Guo2018}.

\begin{figure}[ht!]
\centering
\includegraphics[scale=0.55]{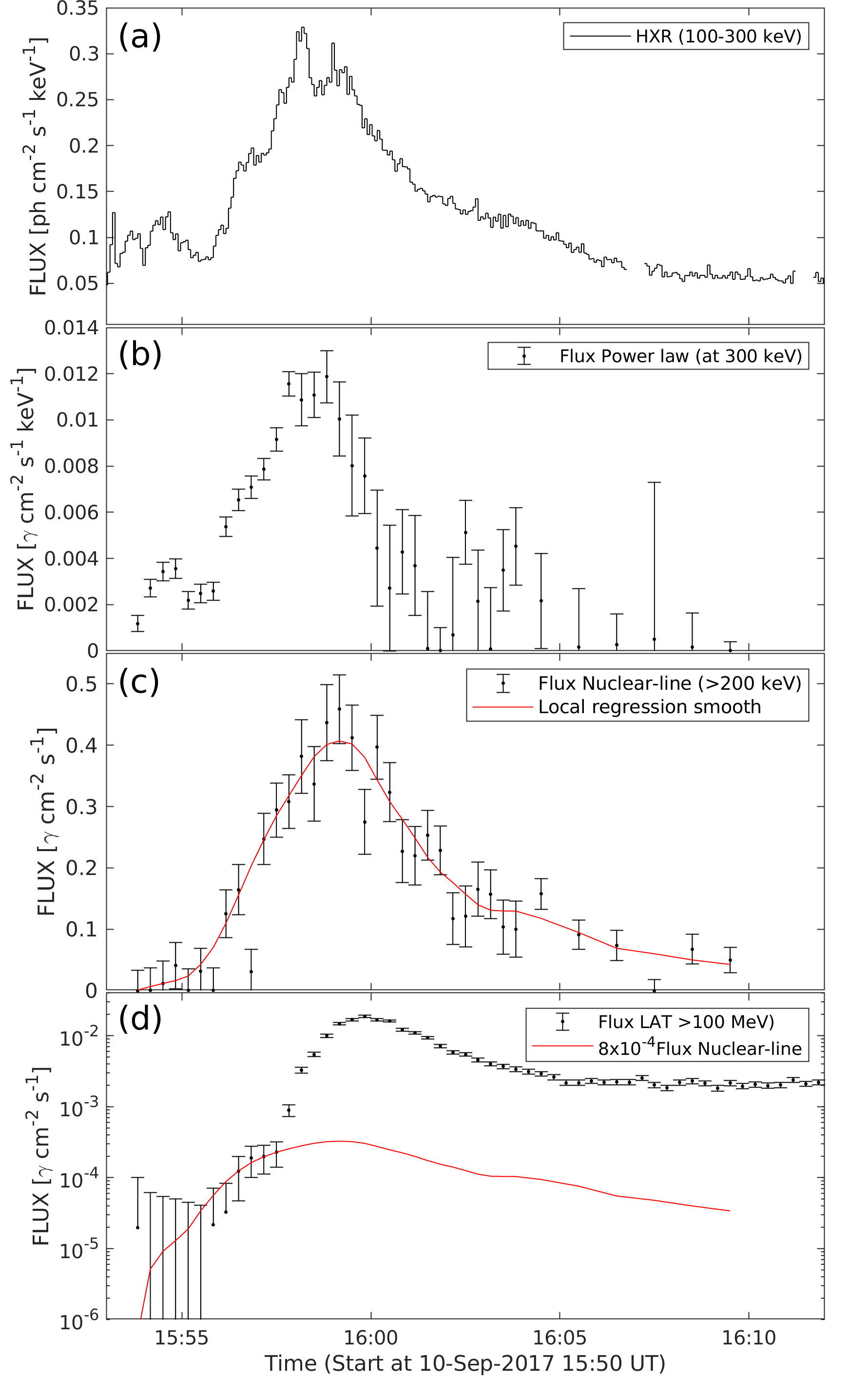}
\caption{From top to bottom: Time history of the 100\,--\,300 keV hard X-ray emission derived from RHESSI measurements (panel a), the flux at 300~keV of a power-law fit to $>$300~keV Fermi/GBM data (panel b), the $>$200~keV flux in narrow and broad nuclear deexcitation lines (panel c), and the $>$100~MeV $\gamma$-ray flux (panel d), during the 2017 September 10 event. \label{fig:FLATLC_b} }
\end{figure}

\subsection{$\gamma$-ray Observations}

Fermi-LAT measured $>$100~MeV $\gamma$-rays for almost 12~hours during and following the flare. The peak flux during this event was the largest recorded by LAT from the Sun since its launch. It was the second longest event recorded so far by Fermi. Figure~\ref{fig:FLATLC} presents the flux time history of the $>$100~MeV $\gamma$-ray emission from the Sun (including celestial and solar quiescent background from within 10$^\circ$ of the sun for plotting purposes) for a two-day period of Fermi-LAT observations (starting on September 10 at 06:00~UT, top panel). The discrete LAT observing intervals with significant solar exposure were typically about 20\,--\,40 minutes in duration every $\sim$90-minute FERMI orbit during the observation. The peak LAT solar exposure varied by a factor of two on alternating orbits and was at the high level during the flare orbit. 

The $>$100 MeV fluxes during the event were derived by fitting background-subtracted LAT spectra with pion-decay spectral templates derived for an isotropic proton distribution assuming a source at a heliocentric angle of 90 degrees. Such an angular distribution is consistent with both fan beam and downward isotropic proton distributions at this heliocentric angle. The detector response matrix used in each fit was appropriate for the solar exposure in each time interval. Due to the high flux of hard X-rays that affected the anti-coincidence system during the first two orbits, we used the special LAT solar impulsive data class in our analysis. Our measured fluxes agree to within 30\% with those given in Table 1 of \cite{Omodei2018} except from 17:33-17:58 UT (the second orbit) where they used the standard LAT data products which had dead-time effects. This explains why our flux at that time is 50\% higher. The bottom panel of this figure displays the fluxes at 20-second cadence during the first orbit (blue shaded area in the top panel) of the event. 



The overall time history of the $>$100~MeV emission is complex. As plotted in the bottom panel of Figure~\ref{fig:FLATLC}, the emission began at 15:56:30~UT and increased slowly until about 15:57:50~UT when it began to increase much more rapidly, reaching a peak near 16:00~UT. We show this early temporal evolution in more detail in Figure~\ref{fig:FLATLC_b} where we plot four signatures of this limb flare: the 100\,--\,300~keV hard X-ray emission derived from RHESSI measurements (panel a), the flux at 300~keV of a power-law fit to $>$300~keV Fermi/GBM data (panel b), the $>$200~keV flux in narrow and broad nuclear deexcitation lines (panel c), and the $>$100~MeV $\gamma$-ray flux (panel d). The time histories shown in panels a) and b) reflect the time histories of a few hundred to several hundred keV flare-accelerated electrons. It is clear that the observed impulsive flare emission at energies $>$300~keV by GBM (panel b) ends earlier than the 100--300~keV hard X-ray emission by RHESSI (panel a), so that the acceleration of the several hundred keV flare-accelerated electrons ends earlier than the few hundred keV electrons. The flux of the $\sim$5 to 30~MeV protons that produce the nuclear deexcitation-line  $\gamma$-rays \citep[][and references within]{Murphy2009} has a time history similar to that of the high-energy electrons, but with a small delay.

In contrast, the $>$100~MeV $\gamma$-ray emission prior to 16:05~UT appears to be comprised of two components: a weak impulsive flare phase visible up to 15:57:30~UT (see the red curve which depicts the nuclear-line time history shown in panel c), normalized to the $>$100~MeV emission prior to 15:57:30~UT) and a much stronger component that begins about a minute after the nuclear-line $\gamma$-rays begin and extends for a much longer period of time. Such late-phase $>$100~MeV $\gamma$-ray emission that is coincident with but temporally distinct from the impulsive-phase of the flare has been observed before; e.g. the events of 2011 September 6 and 24, and 2012 March 5 and 7 \citep{Share2018}. This suggests the presence of an acceleration process that is different from that producing the impulsive-phase emission but that can occur in close temporal proximity to it. 

\begin{figure*}[ht!]
\centering
\includegraphics[scale=0.95]{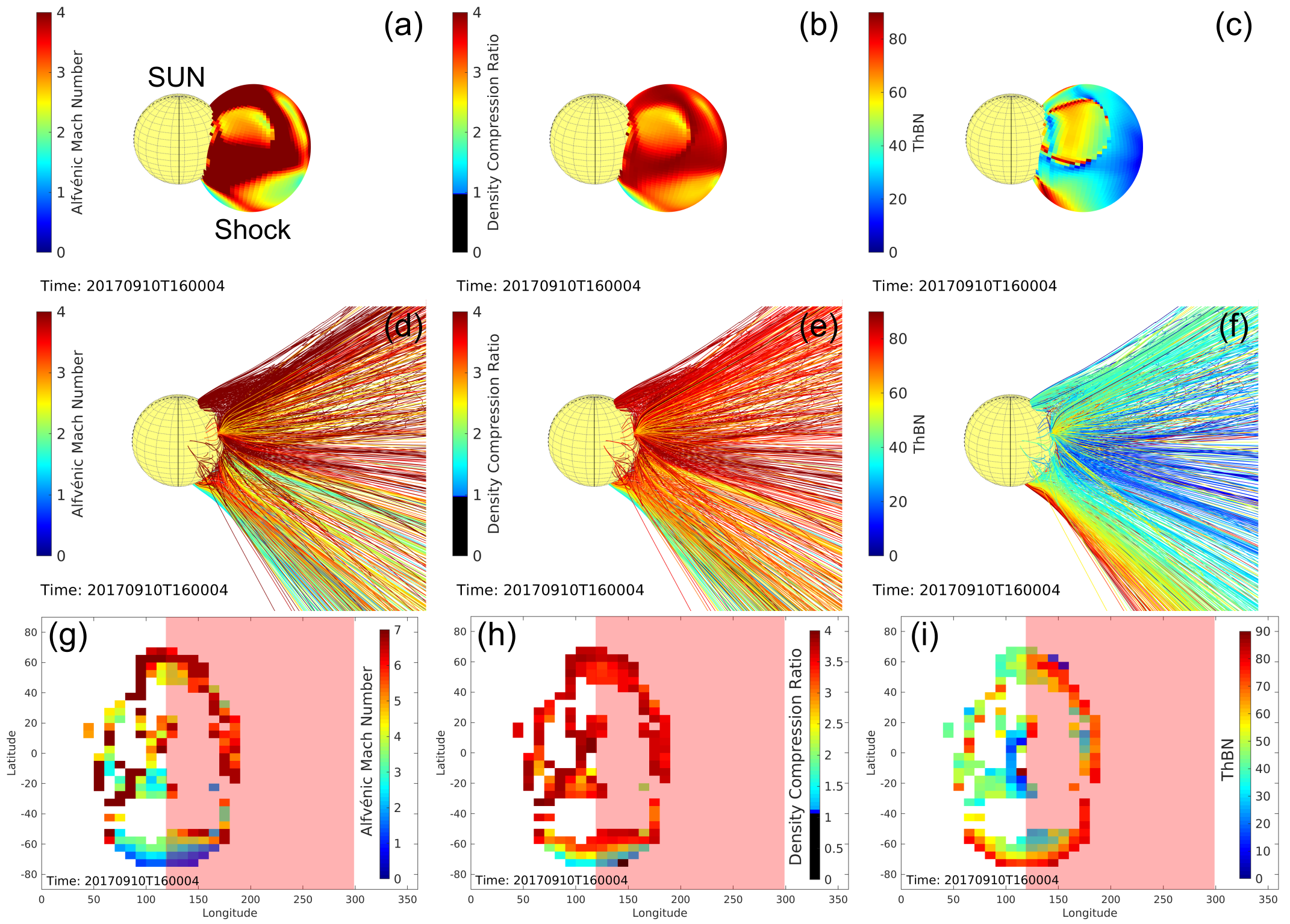}
\caption{Panels (a) to (c): 3D view of the reconstructed pressure wave and the distributions of the modeled shock parameters along the front surface for the 2017 September 10 event at 16:00~UT. From left to right the panels shows the 3D distributions of the Alfv\'enic Mach number ($\mathrm{M_A}$), the density compression ratio ($\mathrm{X}$), and the magnetic field obliquity with respect to the shock normal ($\mathrm{\theta_{BN}}$). Panels (d) to (f): 3D view of the magnetic field lines that connect to the pressure wave. The color of each field line depicts the shock parameter at the point of intersection of the field line with the wave surface. For display purposes only a fraction of the total field lines that are connected to the wave surface are plotted here. Panels (g) to (i): binned latitude\,--\,Carrington-longitude maps, at the solar surface, showing the shock parameters registered to the magnetic field lines connected to the modeled shock wave surface. The pink shaded area marks the part of the solar disk not visible from Earth. \label{fig:3DShockProp} }
\end{figure*}

Following the late phase emission peak near 16:00 UT, the $>$100~MeV flux fell smoothly until about 16:06 UT where it leveled off and began to gradually increase again until the end of the first LAT solar exposure (bottom panel, Figure~\ref{fig:FLATLC}). This increase appears to be a second stage of LPGRE. What is puzzling is that, although the $>$100~MeV flux appears to be falling from a peak during the following solar exposure (top panel, Figure~\ref{fig:FLATLC}), there appears to a factor of about three discontinuity in flux. This led \cite{Omodei2018} to conclude that there is yet another stage of LPGRE that peaked between the two solar exposures. There is an alternative reason for the discontinuity. If the LPGRE is due to particles accelerated at the CME shock, it is possible that a change in the magnetic configuration prevented a large fraction of protons to return to impact the Sun \citep[e.g.][]{Hudson2018}. We study this possibility in Section~\ref{sec:MagnetRatio}. 

\section{ Shock wave reconstruction, 3D Modeling, and Magnetic Field Connections } \label{sec:ShockModel}

The evolution of the CME and its associated pressure wave was well observed by STEREO and near-Earth spacecraft. From the viewpoint of STEREO-A the eruption occurred on the far side of the Sun, well behind its east limb. However, the Extreme UltraViolet Imager (EUVI), and the two SECCHI \citep{Howard2008} coronagraphs COR1 and COR2 recorded with very good coverage the later phase of the eruption (from 1.4 to 15~$\mathrm{R_\sun}$). Additionally, the EUV observations from the Atmospheric Imaging Assembly \citep[AIA:][]{Lemen2012} on board the Solar Dynamic Observatory (SDO) and the WL observations from the Large Angle and Spectrometric Coronagraph \citep[LASCO:][]{Brueckner1995} on board the Solar and Heliospheric Observatory (SOHO) allowed a detailed analysis of the CME and the shock evolution from the initiation point to $\gtrsim$30~R$_\sun$. To extend the shock modeling even higher we used observations of the heliospheric imagers on-board STEREO \citep[HI1 and HI2:][]{Eyles2009} which are designed to observe transients in the IP medium \citep[e.g.][]{Rouillard2011,Rouillard2012}. The shock modeling over this extended time interval is expected to be less accurate than at earlier times.

To model the shock wave parameters in 3D we use the method of \cite{Rouillard2016} and \cite{Plotnikov2017}. We also consider the extensions presented in \cite{Kouloumvakos2019}. Those methods essentially combine the shock reconstructions from imaging to precisely localize the shock wave in 3D and calculate the shock speed along the entire wave front. We start the 3D shock fittings from the low corona using EUV observations (AIA: 1.0\,--\,1.4~$\mathrm{R_\sun}$, EUVI: 1.0\,--\,1.7~$\mathrm{R_\sun}$) and continue with WL coronagraphic observations until the shock wave reaches the outermost extent of the LASCO/C3 FOV (30~$\mathrm{R_\sun}$). From the 3D reconstruction, we calculate the shock wave speed in 3D. The kinematic analysis of the shock wave reveals a very fast evolution shortly after its initiation (estimated to be around 15:53~UT). We find that the shock wave reached a maximum speed at around 16:15 to 16:18~UT. At the shock apex we find a maximum speed of $\mathrm{\sim3600\pm100~km\,s^{-1}}$, while for the flanks, we find a maximum speed of $\mathrm{\sim2700\pm50~km\,s^{-1}}$. 

In a second step, static MHD simulations of the background solar corona and solar wind are used to infer the necessary macroscopic parameters at the shock surface. Compared to full MHD solutions \citep[e.g.][]{Jin2018}, the static simulations do not include the dynamic evolution of the CME and its effect on the post-shock coronal plasma, however the properties of the modeled shock wave can be comparable. For this study, we used the Magneto-Hydrodynamic Around a Sphere Thermodynamic (MAST) model \citep{Lionello2009,Riley2011}. MAST is an MHD model developed by Predictive Sciences Inc. that makes use of the photospheric magnetograms from SDO/HMI as the inner boundary condition of the magnetic field and includes detailed thermodynamics with realistic energy equations accounting for thermal conduction parallel to the magnetic field, radiative losses, and parameterized coronal heating. Using the plasma and magnetic field properties provided from the MAST model of Carrington rotation 2194 (CR2194: starting from 2017-08-16 11:13, ends at 2017-09-12 17:15) and shock reconstructions, we determine in 3-D and along the entire shock front the upstream shock Mach numbers (Alfv\'enic $\mathrm{M_A}$ and fast magnetosonic $\mathrm{M_{fm}}$), density compression ratio ($\mathrm{X}$), and the shock geometry $\mathrm{\theta_{BN}}$ which is defined as the magnetic field obliquity with respect to the shock normal \cite[see][for a detailed description of the techniques]{Kouloumvakos2019}. This modeling is performed with a temporal resolution of one minute, from the low corona to IP space at $\sim$150~$\mathrm{R_\sun}$ (i.e. $\sim$0.6~AU). The \cite{Kouloumvakos2019} and \cite{Plotnikov2017} modelling reveals that shock waves have significant event to event variability and shock properties can vary significantly along a single shock surface. We note that the  Mach number is related to the energy being processed by the shock, and it is hence an indicator of the strength of the shock \citep[][]{Burgess2015}. The compression ratios increase with increasing Mach number, and they are also commonly used as proxies for the shock strength. 

To infer magnetic connections between the shock and either the solar disk or a point of in situ measurement, we perform field-line tracing (FLT) using the magnetic field vector data from the MAST 3D cubes. We note, however, that weaknesses of using the MAST data to trace magnetic field lines are low grid resolutions $\mathrm{>10~R_\sun}$ and numerical diffusion. These have a minimal effect on our estimates since for this work we consider a large amount of field lines to perform the statistics over the visible disk. We start the FLT at the wave front and continue sunward or antisunward until one of the boundaries (inner or outer) of the MAST simulation model is reached. In this paper, we consider both open- and closed-field lines that intersect the reconstructed shock surface. The FLT is performed at every time step between launch and until $\sim$12~hours into the event. We register for each field line the shock parameters determined at the intersection point of the field line with the shock surface. 

We plot in Figure~\ref{fig:3DShockProp} an example of the information on the shock parameters (panels a to c) and the magnetic field connections to the solar disk (panels d to f) at 16:00 UT. We depict in different panels the 3D distribution of the modeled parameters and the magnetic field lines that thread the shock surface. As can be seen in panel (a) and (b), there are multiple regions along the 3D reconstructed pressure wave front where a strong shock (i.e. $\mathrm{M_A}>3$ and $\mathrm{X}\sim4$) has formed by 16:00~UT and that are magnetically connected to the visible disk (see panel d and e). Those regions are located mainly at the northern and southeastern flanks of the pressure wave, as well as at its apex; and efficient shock acceleration is expected to take place at those locations because the shock wave is super-critical ($\mathrm{M_A>M_c}$). For collisionless shock waves there is a critical Mach number, $\mathrm{M_c}$, above which simple resistivity cannot provide the total shock dissipation. When $\mathrm{M_A > M_c}$, e.g. in the super-critical case, a significant part of upstream ions are reflected on the shock front gaining an amount of energy that enables them to be injected into the acceleration process. In the sub-critical case the ions are not reflected and this significantly reduces the acceleration efficiency. $\mathrm{M_c}$ mostly depends on local plasma conditions (plasma $\mathrm{\beta}$) and shock geometry $\mathrm{\theta_{BN}}$. For $\mathrm{\beta<1}$, the $\mathrm{M_c}$ varies from $\sim$1.53 for parallel and $\sim$2.76 for perpendicular shocks.

At the shock apex we find that the geometry is mainly quasi-parallel (panel c and f). There are extended locations at the eastern and southeastern flanks where the shock is mainly quasi-perpendicular. Efficient particle acceleration is usually considered to take place at a quasi-perpendicular geometry if the upstream magnetic field is sufficiently turbulent. LOFAR radio imaging observations reveal that bright Type-II radio emission that originates at both the southern and northern flanks of the white-light shock that was driven by the CME \cite[see Figure 2 of][]{Morosan2019} for the same time interval we show in Figure~\ref{fig:3DShockProp}.

\begin{figure*}[ht!]
\centering
\includegraphics[scale=0.56]{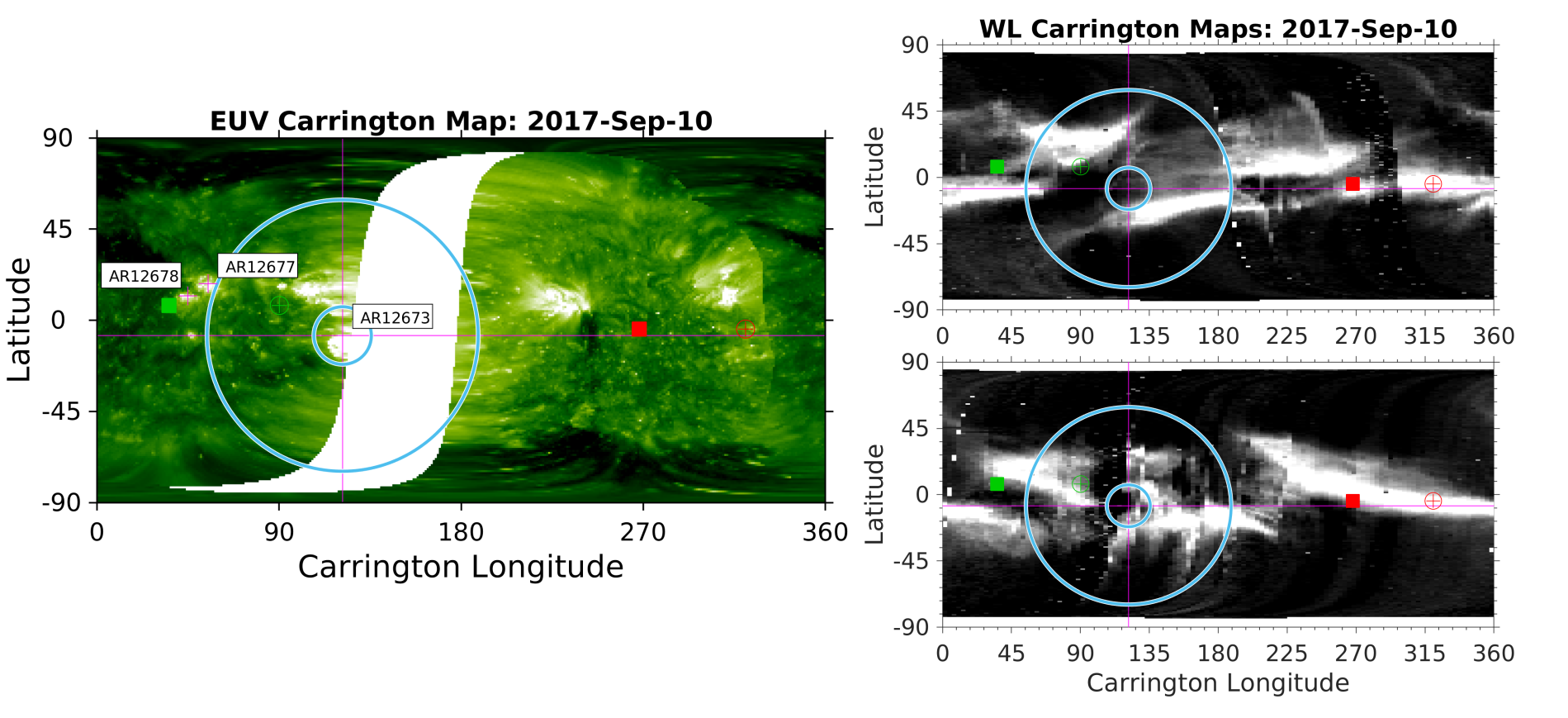}
\caption{Carrington maps in EUV from AIA and STA/EUVI observations (left panel) and of the white-light corona at an altitude of 3~R$_\sun$ over the east and west limb of the Sun from LASCO/C2 (right panels) observations (top and bottom panels, respectively). The overplotted square points depict the position of the Earth (green) and STEREO-A (red) while the circles denote the position of their magnetic connections at 2.5~R$_\odot$. The location of AR12673, where the event takes place, is depicted by the intersection of the two thin red lines. On the EUV maps, we also show the location of AR12677 and AR12678. The two light blue concentric cycles show the change of the angular extent of the shock flanks during the first hour of the expansion. \label{fig:Corona_cond} }
\end{figure*}

During the early evolution of the shock wave in the low corona and nearly after 16:00~UT, our modeling reveals that there are regions on the northeast/east flank and towards the edges of the shock wave at the solar surface, that lose strength (i.e. $M_A$ and X drop). This happens when the shock front reached a cluster of two ARs (AR12677 and AR12678) visible in the EUV Carrington map of Figure~\ref{fig:Corona_cond}. The two ARs are located northeast/east of the eruption and close to disk center. The wave propagation was interrupted by those regions and could not be traced towards the eastern limb. The shock is more likely to progressively degenerate into a fast mode wave rather than to continue as a shock. This can been seen in the EUV observations of SDO/AIA during the event \citep[also see][]{Liu2018,Hu2019}. At low coronal heights, the high Alfv\'en speed at the nearby ARs (AR12677 and AR12678) certainly affects the evolution of the shock, at least during the early expansion phase (the first $\sim$20 minutes). Similar conditions seem to apply for regions near the southern flank. When the shock has evolved higher in the corona, i.e. after 16:30~UT, the lateral expansion of the shock flanks becomes more important. The Mach number progressively increases because the Alfv\'en speed drops faster than the shock wave decelerates during its expansion to higher altitude.


In Figure~\ref{fig:Corona_cond} (right panels) we show Carrington maps of the white-light corona, constructed by using SOHO/LASCO white-light C2 images from 2017 August to September, at an altitude of 3~R$_\odot$ over the east (top map) and west (bottom map) limb of the Sun. These maps are constructed by extracting the WL intensity using a circular slice at the heliocentric radial distance of 2.5~R$_\odot$ at SOHO/LASCO white-light C2 images. For each pixel we calculate the elongation and position angle and map them in a Carrington-longitude versus latitude map. This procedure is repeated for each LASCO-C2 image during 2017 August and September. The WL Carrington maps give a better perspective of the coronal configuration during the shock expansion in the high corona. The streamer belt associated with the neutral current sheet dominates the general configuration of the corona. There are also other features such as additional plasma sheets that do not correspond to the source surface neutral line and transient features like CMEs or dark voids that correspond to coronal holes. All of the above features, except the coronal hole areas, appear bright in the maps. 

During the shock wave's lateral expansion, the shock flanks pass through streamer regions that are located east of the AR12673. In white light SOHO/LASCO images, the deflection of streamers is also clearly observed. Streamers have typically low Alfv\'en speed, high plasma density, and weak magnetic field. Additionally, at those regions, the solar wind speed is lower. As a consequence of the lower characteristic speed of the medium though which the shock wave propagates, the shock strength (Mach number) is further increased so the compression ratio of the shock wave can be much enhanced. Additionally, at those regions the quickly varying shock geometry (from quasi-perpendicular to quasi-parallel) and the natural trapping of particles at closed magnetic fields could lead to very efficient particle acceleration \citep[e.g.][]{Kong2019}.

To better exploit the results of the shock wave modeling and perform an analysis of the spatial and temporal variability of shock parameters at the field lines connected to the visible solar disk, we construct latitude\,--\,longitude maps, similar to \cite{Plotnikov2017}, considering both the open and closed field lines. In Figure~\ref{fig:3DShockProp} (panels g to i) we show an example of the constructed maps in a one-minute interval at 16:00 UT for the shock parameters considered in this study. We plot the parameters versus Carrington longitude, and with the pink shaded area we mark the part of the solar disk which was not visible from Earth. The maps have been constructed using a pixel size of 5$^\circ$x10$^\circ$ in latitude\,--\,longitude and for each pixel we compute the mean of the shock parameters. 

The colored pixels in the latitude\,--\,longitude maps show the magnitudes of the different shock parameters projected onto the solar surface at the footpoints of the magnetic field lines. A significant fraction of the field lines intersected by the shock flanks at this time are closed, with footpoints both near the active region and at more distant locations. The elliptical shape of these locations indicates that it was parameters of the flank of the shock that were mostly projected onto the solar disk. This is reflected in generally quasi-perpendicular geometry of the shocks crossing the field lines. Strong shock regions were connected to the visible disk but also to sites beyond the limb of the Sun (pink shaded region). We use the magnitudes of the shock parameters in similar maps produced at different times to compare with the observed time history of gamma-ray emission. We do this in the next section.

\section{ESTABLISHING THE SHOCK ORIGIN OF LATE PHASE $>$100 MeV GAMMA RAY EMISSION} \label{sec:GrayShock}

In this section we examine the connection of the temporal variability of the modeled shock parameters averaged over the regions that are magnetically connected to the visible disk with the observed $\gamma$-ray emission. If the expanding shock wave plays a significant role in the acceleration of those protons that interact in the solar chromosphere and photosphere, we expect that the time variation of the parameters should follow the time profile of the $>$100~MeV $\gamma$-ray flux.

In Section~\ref{sec:4p1} we discuss the temporal evolution of the shock parameters magnetically connected to the visible solar disk during the first orbit of LAT observations, that is dominated by what appears to be a first stage of LPGRE followed by the onset of the second stage of emission. In Section~\ref{sec:4p2} we discuss the temporal evolution of the parameters at later times in the event. We recognize that these parameters do not directly reflect the flux of protons reaching the solar atmosphere where they can interact. Therefore in Section~\ref{sec:MagnetRatio} we create an empirical function that better reflects how the shock relates to the accelerated proton flux and that also estimates the efficiency for the protons to reach the solar atmosphere in the presence of magnetic mirroring. We find that introducing this function significantly improves the agreement between the shock estimates of the proton flux and the $\gamma$-ray observations. We show that the shock model clearly explains what appears to be two stages of LPGRE.

\subsection{ Early temporal evolution of $\gamma$-rays and shock parameters magnetically linked to the Sun} \label{sec:4p1}

\begin{figure}[ht!]
\centering
\includegraphics[scale=0.44]{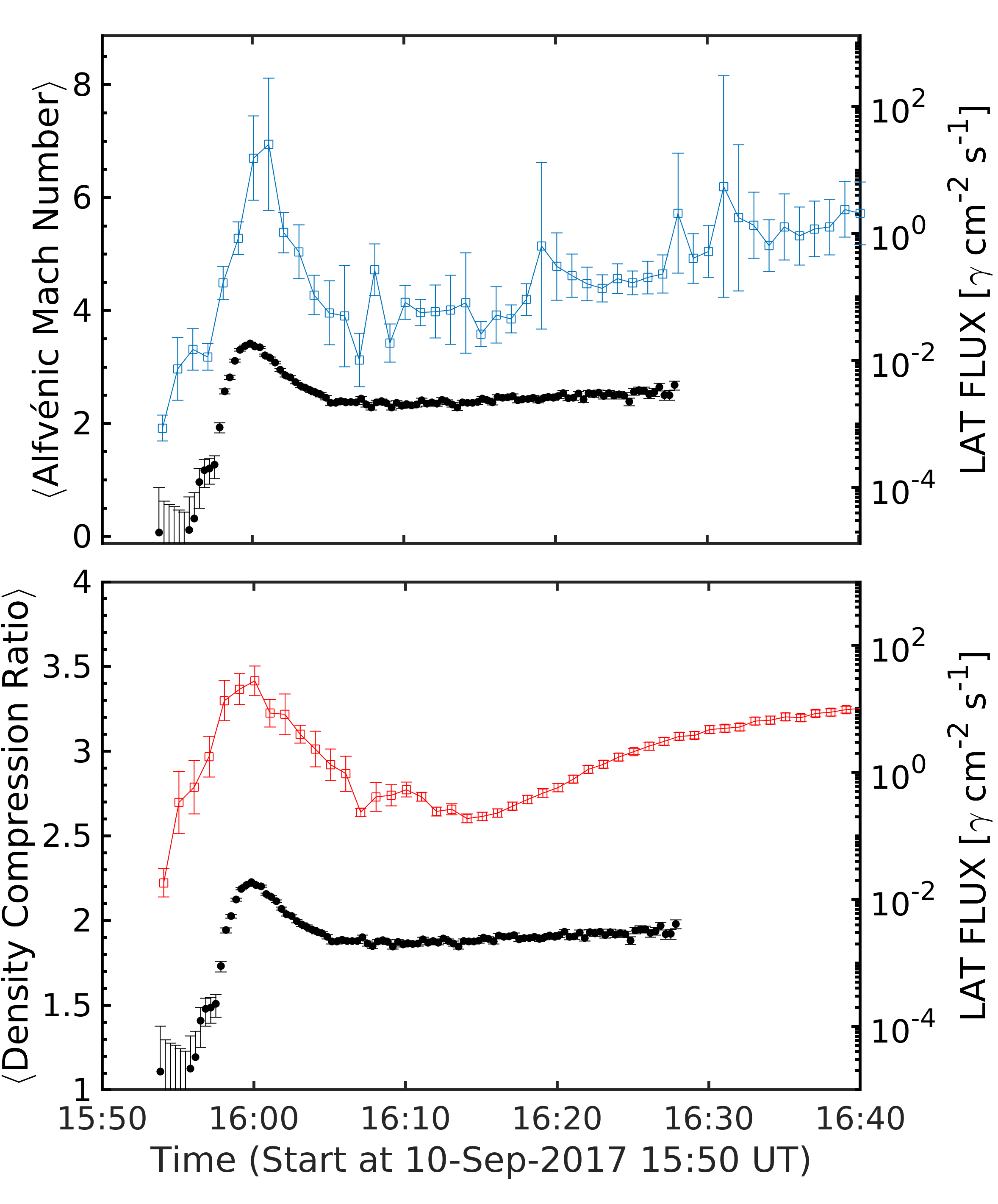}
\caption{Comparison between the evolution of the $>$100~MeV $\gamma$-ray emission from LAT (black-filled data points in both panels) and the mean values and uncertainties of two shock parameters, from 15:50~UT to 16:40~UT. Top panel: The mean Alfv\'enic Mach number ($\mathrm{\langle M_{A} \rangle}$) for the field lines connected to the visible disk (blue connected data points with errors). Bottom: Similar plot for the evolution of the mean density compression ratio ($\mathrm{\langle X \rangle}$, red connected data points with errors).  \label{fig:GRAYSShockPMAX} }
\end{figure}

Figure~\ref{fig:GRAYSShockPMAX} presents a comparison of the time history of the $>$100~MeV $\gamma$-ray emission from the start of the event until the end of the first FERMI-LAT observation interval together with the evolution of the mean Alfv\'enic Mach number (upper panel) and density compression ratio (bottom panel) for the magnetic field lines returning to the visible solar disk. The vertical bars of the shock parameters depict the 3$\sigma$ uncertainties of the evaluated shock parameters. As it can be seen, a striking similarity exists between the temporal evolution of the simulated shock parameters and the evolution of the observed $\gamma$-ray emission. 

At the start of the $\gamma$-ray event we find that the eastern flank of the shock is already magnetically connected to the visible disk. This is expected because the CME erupts in the vicinity of the west limb of the Sun as viewed from Earth. Additionally, the regions of the shock that are magnetically connected to the visible disk become, on average, super-critical ($\mathrm{M_A>M_c}$) at around 15:55~UT. Because the shock is on average quasi-perpendicular during its early evolution, we have that $\mathrm{M_c\sim2.76}$. The early temporal evolution of the Alfv\'enic Mach number after the time that the shock becomes, on average, super-critical follows the early time history of $\gamma$-rays that we previously attributed to impulsive phase flare emission. We cannot conclude if this is coincidental; if not, this would suggest the possibility that there may be three stages of LPGRE related to shock acceleration. Two minutes later at about 15:57:40 UT, the $>$100~MeV $\gamma$-ray emission increased rapidly marking what we assumed to be the first stage of the late-phase emission. As is clear from Figure~\ref{fig:GRAYSShockPMAX}, both the mean Alfv\'enic Mach number and the mean density compression ratio increase rapidly until 16:00~UT.

At 16:00~UT, the shock wave regions that were connected to the visible disk reached their maximum strength. This is reflected in both the mean Alfv\'enic Mach number and the mean density compression ratio that reach maxima within one minute of each other. This is concomitant with the most intense peak in $>$100~MeV $\gamma$-ray emission at 15:59:50 UT. \cite{Jin2018} have also showed a similar relation between the evolution of the shock compression ratio and the $>$100~MeV $\gamma$-ray emission for the 2014 September 01 event. Remarkably, both the $\gamma$-ray flux and the shock parameters also follow one another closely after the peak. Both decrease until 16:05 UT when a minimum is reached and what appears to be a new stage begins. There are some finer time structures visible in the shock parameters that are not apparent in the $\gamma$-ray time history. In addition, the parameters increase more rapidly than the $\gamma$-ray flux. This suggests that another parameter may affect the number of protons reaching the solar atmosphere at this time. We will return to this aspect in Section~\ref{sec:MagnetRatio} where we examine the effects of magnetic mirroring.

\begin{figure}[ht!]
\centering
\includegraphics[scale=0.45]{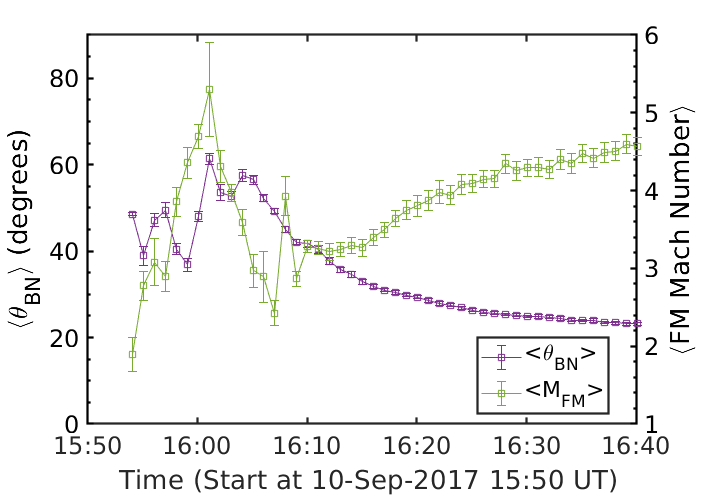}
\caption{Evolution of the mean shock geometry, $\mathrm{\langle \theta_{BN} \rangle}$ (purple points), and the mean fast magnetosonic Mach number, $\langle \mathrm{M_{fm}} \rangle$ (green points), over the same time interval as Figure~\ref{fig:GRAYSShockPMAX}. \label{fig:GRAYS2ndProp} }
\end{figure}

In Figure~\ref{fig:GRAYS2ndProp} we examine the evolution of the means of other shock parameters: the fast magnetosonic Mach number ($\mathrm{M_{fm}}$) and the shock geometry, $\mathrm{\theta_{BN}}$. Because $\mathrm{M_{fm}}$ depends on both the Alfv\'en speed and the sound speed, its use instead of the Alfv\'enic Mach number serves as a further test of the observed variability since we avoid excessively high Mach number values in coronal neutral regions where the magnetic field intensity drops to very low levels. From Figure~\ref{fig:GRAYS2ndProp} we see that the temporal evolution of $\mathrm{M_{fm}}$ is similar to those of $\mathrm{M_{A}}$ and $\mathrm{X}$ that are presented in Figure~\ref{fig:GRAYSShockPMAX}.

From the study of $\mathrm{\theta_{BN}}$ in Figure~\ref{fig:GRAYS2ndProp} we find that the shock crossed field lines connected to the visible disk on average at an oblique angle, from the start of the event until $\sim$16:01~UT. After that time, the shock moved closer to a quasi-perpendicular geometry for $\sim$5~minutes. After 16:10~UT, $\mathrm{\theta_{BN}}$ progressively changed from an oblique to quasi-parallel geometry, asymptotically approaching $20^\circ$ by 16:50 UT. Simulations of diffusive shock acceleration have demonstrated that high-energy particles can be accelerated efficiently in both quasi-parallel and quasi-perpendicular shock waves \citep[e.g.][]{Zank2006,Vainio2014,Afanasiev2018}. Overall, at quasi-perpendicular shocks the acceleration time is very short compared to parallel shocks and a suprathermal seed population is needed for the acceleration to start. At quasi-parallel shocks the particle acceleration time is longer but the acceleration can start even with thermal-energy particles. Additionally, \citet{Zank2006} demonstrated that the maximum energy achieved at quasi-perpendicular shocks is less than that at quasi-parallel shocks near the Sun, and the reverse is true farther from Sun (see their Figure 7).

Comparing the shock properties from our shock modeling with that of the 2014 September 01 event modeled by \cite{Plotnikov2017} and \cite{Jin2018}, we find that the shock wave regions that were connected to the visible disk become super-critical quickly after the CME initiation and the shock strength is on average higher than the 2014 September 01 event. The CME and the driven shock wave of the 2017 September 10 event is much faster than the CME/shock wave of the 2014 September 01 event, so this is one reason that a stronger shock wave is observed. The evolution of the mean shock geometry for the regions that were connected to the visible disk is similar to the 2014 September 01 event. The shock geometry for the 2017 September 10 event is more oblique than quasi-perpendicular at the early stages of the shock evolution.

\begin{figure}[ht!]
\centering
\includegraphics[scale=0.44]{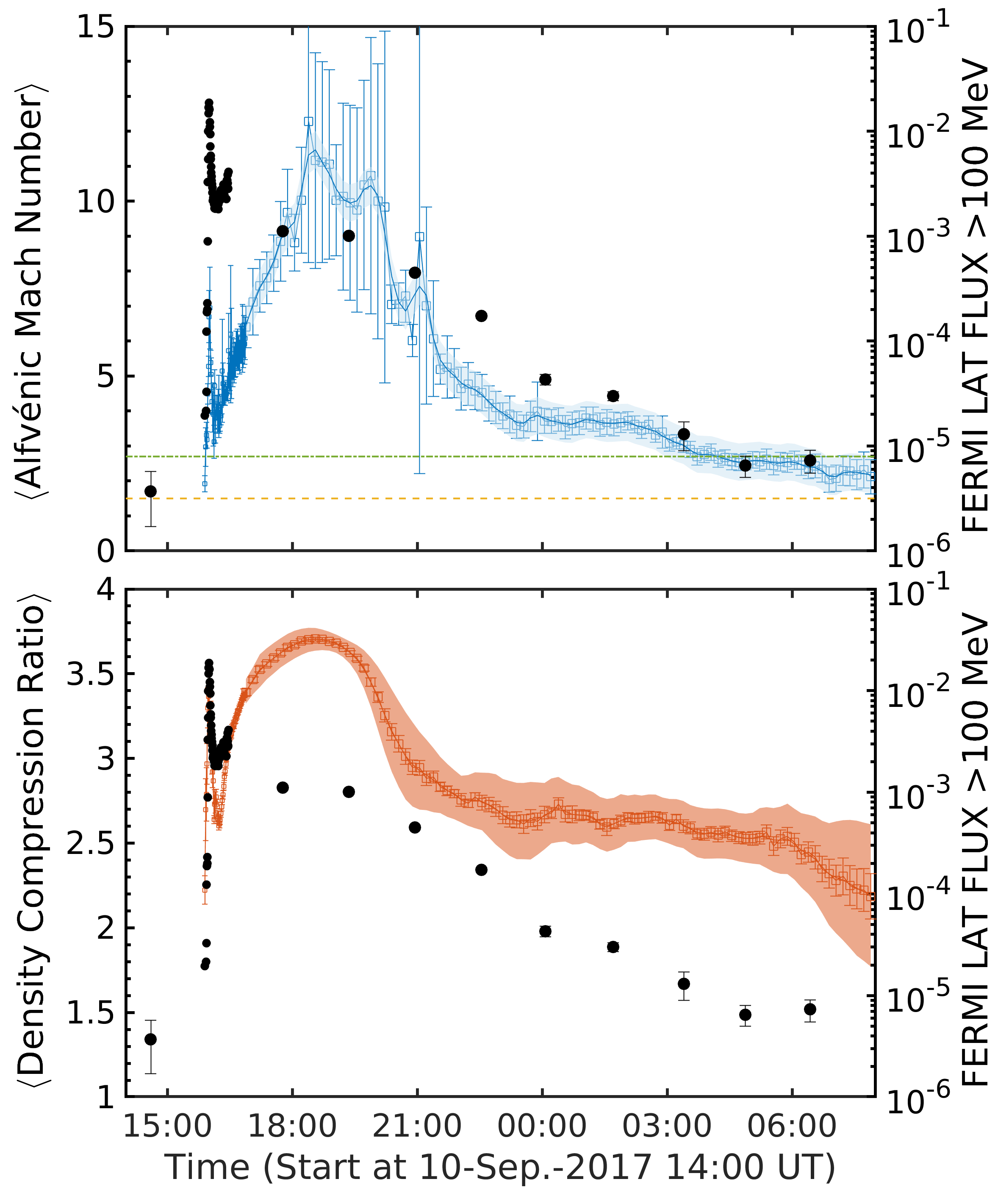}
\caption{Comparison between the time history of $>$100~MeV $\gamma$-ray emission from LAT (black markers in both panels) and the evolution of two shock parameters, during the 2017 September 10 $\gamma$-ray event. Top: Comparison with the evolution of the Alfv\'enic Mach number ($\mathrm{M_{A}}$) for the field lines connected at the visible disk (blue). Bottom: Similar plot for the evolution of the shock density compression ratio (X) (red). The vertical bars on the two shock parameter points depict the 3$\sigma$ uncertainties of the parameters based on statistics and the colored areas depict the estimated additional uncertainty of the parameters due to systematic uncertainties in the extended shock fittings. \label{fig:GRAYSPropExtent} }
\end{figure}

\subsection{Late temporal evolution of $\gamma$-rays and shock parameters magnetically linked to the Sun} \label{sec:4p2}

\begin{figure}[ht!]
\centering
\includegraphics[scale=0.43]{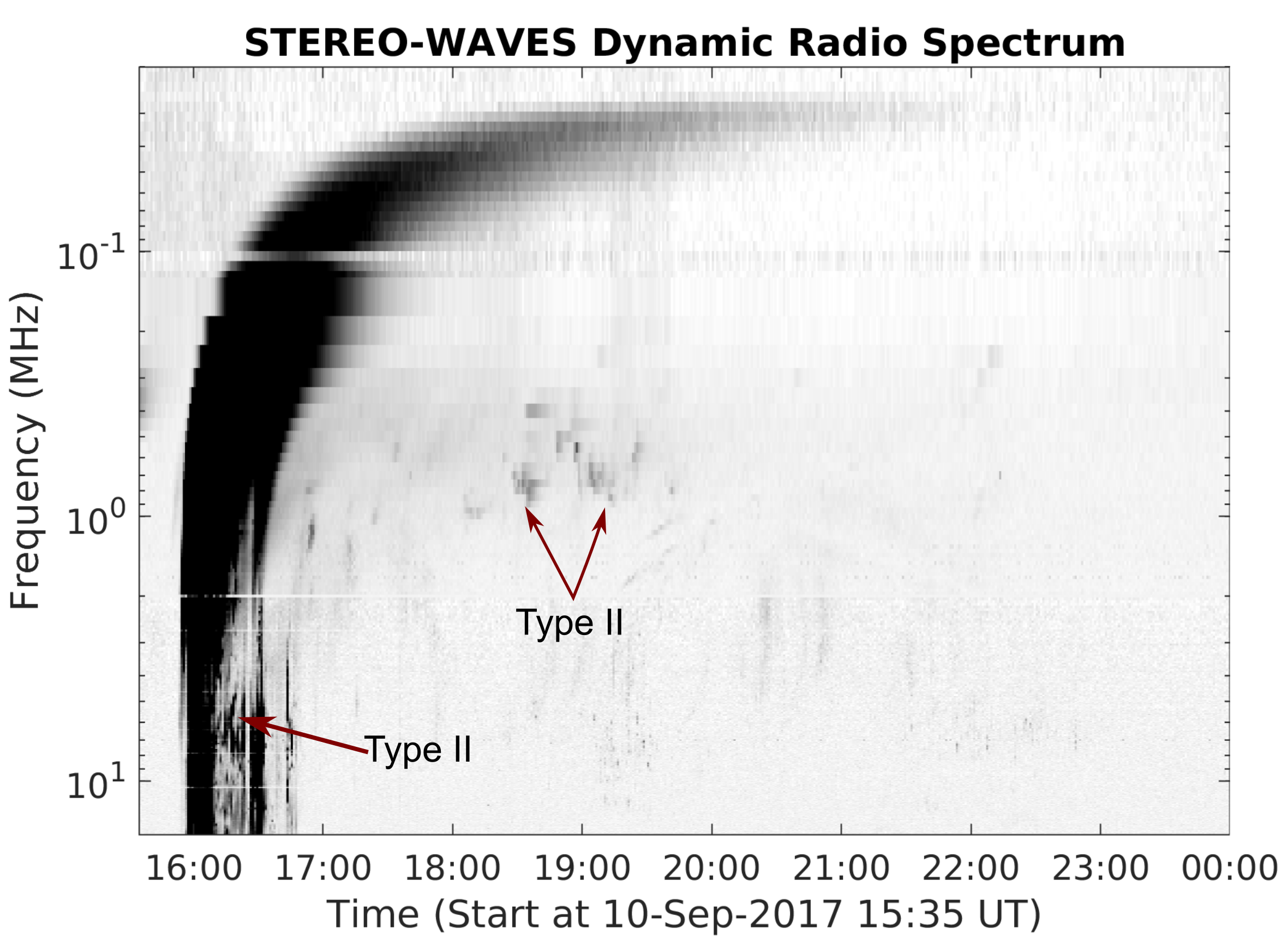}
\caption{Dynamic radio spectrum from STA/WAVES from 2017 September 10 15:35~UT to the end of the day. The regions with enhanced Type-II radio emission are labeled.
\label{fig:RADIOWIND} }
\end{figure}

Below we examine the connection between the shock parameters and the flux of the second stage of the late-phase $\gamma$-ray emission which lasted almost 12 hours. In Figure~\ref{fig:GRAYSPropExtent} we compare the evolution of the modeled shock parameters $\mathrm{M_{A}}$ and $\mathrm{X}$ on field lines returning to the Sun with the LAT $\gamma$-ray emission from the onset, before 16 UT, to the time the emission returns to background at 6 UT on the following day. As can be seen in both Figures~\ref{fig:GRAYSShockPMAX} and~\ref{fig:GRAYSPropExtent}, the $\gamma$-ray emission was rising after the intense peak at 16 UT. From this one might infer that the flux would reach a second maximum in later observations, and such a maximum appears to have been reached between 17:30 and 20:00 UT, the same time that the two plotted shock parameters reached their peak values. Radio observations show that there was also an enhancement of the Type-II radio emission (see later discussion for Figure~\ref{fig:RADIOWIND}) at this time. However, the discontinuity of about a factor of about three in the $\gamma$-ray flux between the first and second LAT solar exposures cannot be explained from the evolution of the shock parameters. It is possible that a change in the magnetic configuration  prevented a large fraction of protons to return to impact the Sun. We examine this in the next Section.

After $\sim$20:00~UT the shock strength (both $\mathrm{M_{A}}$ and X) and the $\gamma$-ray emission both started to decrease, until 05:30 on 2017 September 11 when the $\gamma$-ray flux reached background. The shock is mostly quasi-parallel during this second stage of the late-phase emission, and it is reasonable to assume that $\mathrm{M_c<2.7}$. Therefore, as $\mathrm{M_{A}}$ asymptotically reached a minimum plateau of $\mathrm{\approx2.5}$ after $\sim$06:00~UT, one would expect that particle acceleration would be significantly reduced. We would still expect there to be particle acceleration up until the time when $\mathrm{M_c}$ was less than 1.5.

In Figure~\ref{fig:RADIOWIND} we show the STEREO-A/WAVES dynamic radio spectrum from 15:35~UT to the end of 2017 September 10. Type-II radio emission was observed during both stages of the late-phase emission. From Figure 1 of \cite[][]{Morosan2019}, the brightest shock emission appears to have commenced around 15:59:10~UT at low radio frequencies in the range 30--50 MHz, close to the time when the shock parameters reached their first maximum according to our Figure~\ref{fig:GRAYSShockPMAX}. The shock emission can be clearly traced until 16:30~UT at radio frequencies in the range 7--14 MHz at the STEREO-A/WAVES radio spectrum of Figure~\ref{fig:RADIOWIND}. The shock emission is also intense around 18:00 to 20:00~UT, the time when the shock compression ratio reaches the maximum plateau according to our modeling (see Figure~\ref{fig:GRAYSPropExtent}). The later increase of the Type-II intensity is consistent with the increase of $\mathrm{M_{A}}$, but most important it suggests that the shock front interacts with the various coronal structures such as streamers, as suggested by the analysis of Section~\ref{sec:ShockModel}. For regions at the shock flanks that propagate through a streamer, the shock geometry in this case favors the excitation or enhancement of Type-II radio emission \citep{Kong2015,Frassati2019}. From the emission frequency of the Type-II radio burst at 18:35~UT we find that the source region is located at a heliocentric distance of 16.1$\pm$5.1~R$_\odot$. At the same time this height is more consistent with the location of the shock flanks rather than the shock apex. Overall, the radio measurements also provide a suggestion that there are two stages of particle acceleration that our modeling and the Fermi data indicate.

\subsection{Temporal evolution of $\gamma$-rays and shock-accelerated particles returning to the Sun} \label{sec:MagnetRatio}

In this section, we construct a basic model to estimate the temporal evolution of shock-accelerated particles that can return to the Sun and compare it with the observed $\gamma$-ray emission. In order to do this we construct an empirical quantity that reflects the numbers of shock-accelerated particles that can return to the Sun. The constructed empirical relation has two parts, one that approximates the shock's acceleration, $\epsilon_{sh}$, and one that estimates the fraction of particles that reach the solar atmosphere without being mirrored. The total expression is the product of these two components and a constant $\alpha$, $\mathrm{F' = \alpha f_\eta \epsilon_{sh}}$. For $\epsilon_{sh}$ we assume here that it is given by $\mathrm{(M_A - M_c)^2}$, where $\mathrm{M_A}$ and $\mathrm{M_c}$ are the Alfv\'enic and critical Mach numbers, respectively \citep[see][]{Kouloumvakos2019}. The calculation is performed using the mean shock parameters for the magnetic field lines returning to the visible solar disk. The first term, $\alpha$, is a multiplying constant to visually match the derived $\mathrm{F'}$ with the observed $\gamma$-ray flux. 

We define $\mathrm{f_\eta}$ in an ideal scattering-free environment and assuming that the high-energy protons are distributed isotropically at the shock. The fraction of particles that reach the solar atmosphere without being mirrored by the magnetic force and precipitate is a function $\mathrm{f_\eta = 1-\cos(\alpha_0)}$, where $\sin^2\alpha_0 = 1/\mathrm{\eta}$ and $\mathrm{\eta}$ is the magnetic mirror ratio. The mirror ratio is defined as $\mathrm{B_{R_\odot}/B_{sh}}$, where $\mathrm{B_{sh}}$ is the strength of the magnetic field upstream of the shock and $\mathrm{B_{R_\odot}}$ is the magnetic field at the solar surface \citep[e.g.][]{Klein2018}. We calculate the magnetic mirror ratio for the field lines that are connected to the shock surface and we perform a statistical analysis to derive the evolution of $\mathrm{\langle\eta\rangle}$, similar to the analysis performed for the estimation of the mean shock parameters. 

\begin{figure}[ht!]
\centering
\includegraphics[scale=0.38]{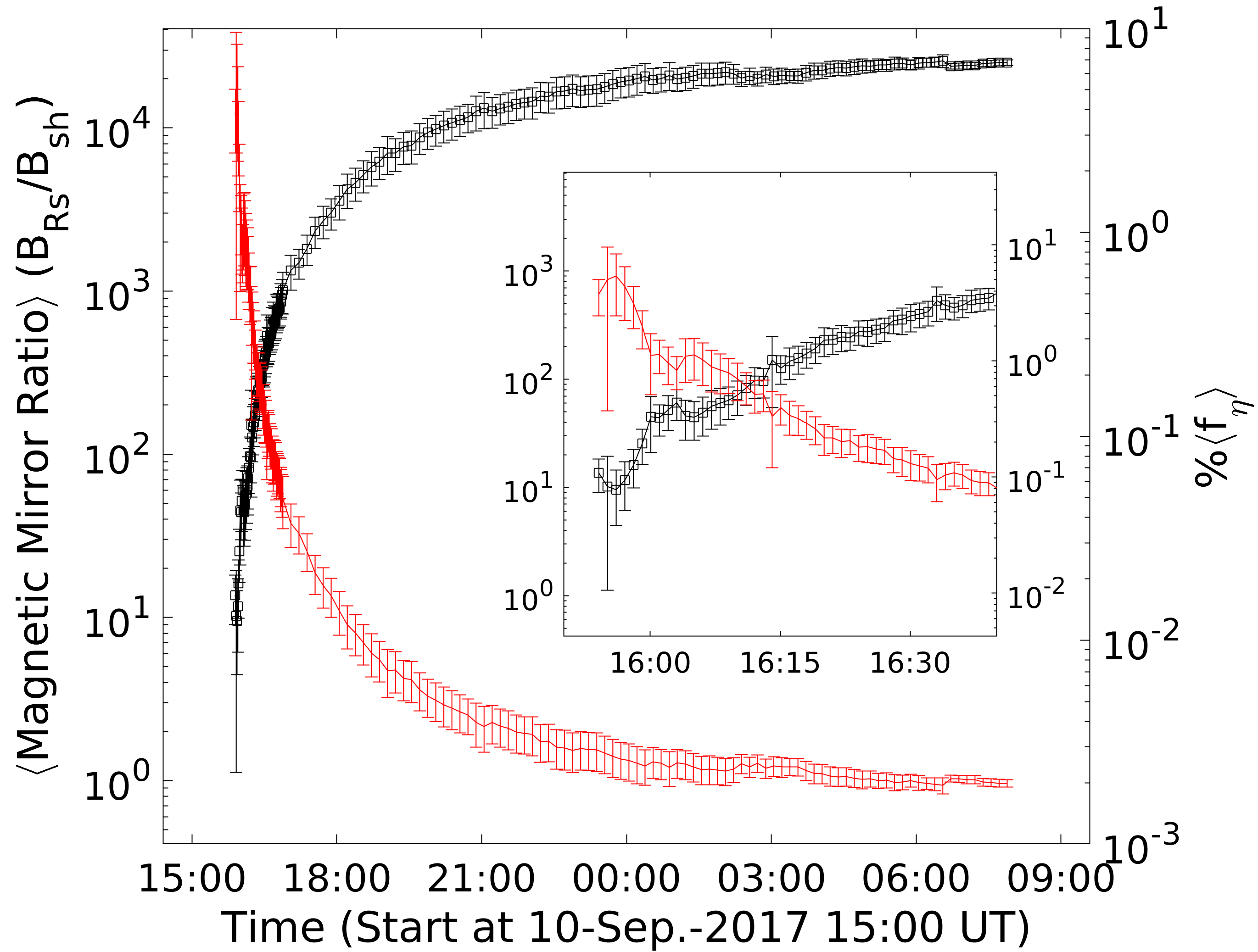}
\caption{Evolution of the mean magnetic mirror ratio (black curve) and the mean fractional  precipitating percentage $\mathrm{\langle f_\eta\rangle\%}$ (red curve), for the field lines connected to the visible disk and the 3$\sigma$ uncertainties based on the statistics. The inset panel presents a zoomed in version during the first stage of the late-phase emission from 15:50~UT to 16:40~UT.
\label{fig:Mratio} }
\end{figure}

In Figure~\ref{fig:Mratio} we show the evolution of $\mathrm{\langle\eta\rangle}$ throughout the LAT observations. For the first interval of LAT observations, we find that $\mathrm{\langle\eta\rangle}$ varies from $\sim$10 near the start to $\sim$400 at the end of the LAT observing interval. Specifically, at 16:00~UT when the $\gamma$-ray emission is maximum $\mathrm{\langle\eta\rangle}$ is $\sim$40. At later intervals $\mathrm{\langle\eta\rangle}$ increases monotonically and asymptotically reaches a plateau at $\sim$2550 near the end of the modeling interval.

Using the $\mathrm{\langle\eta\rangle}$ we estimate the precipitating fraction ($\mathrm{\langle f_\eta\rangle}$), which is also presented in Figure~\ref{fig:Mratio}. $\mathrm{\langle f_\eta\rangle}$ varies from $\sim$5.4\% ($\langle\eta\rangle\sim$9.5) at $\sim$15:56~UT to $\sim$1.1\% ($\langle\eta\rangle\sim$45) at $\sim$16:00~UT. After the maximum of the first stage and until the end of the LAT observations, the fraction of precipitating particles significantly drops from $\mathrm{\sim}$0.055\% ($\langle\eta\rangle\sim$900) at $\sim$16:50~UT to $\mathrm{\sim}$0.002\% ($\langle\eta\rangle\sim$2495) at $\sim$06:03~UT on September 11. As we discuss below, this order-of-magnitude decrease in the percentage of precipitating protons from about 16:30 UT to about 18:00 UT can explain an inconsistency between the relative intensities of $\gamma$-ray fluxes and the shock parameters plotted in Figure~\ref{fig:GRAYSPropExtent} during those times. 

\begin{figure}[ht!]
\centering
\includegraphics[scale=0.33]{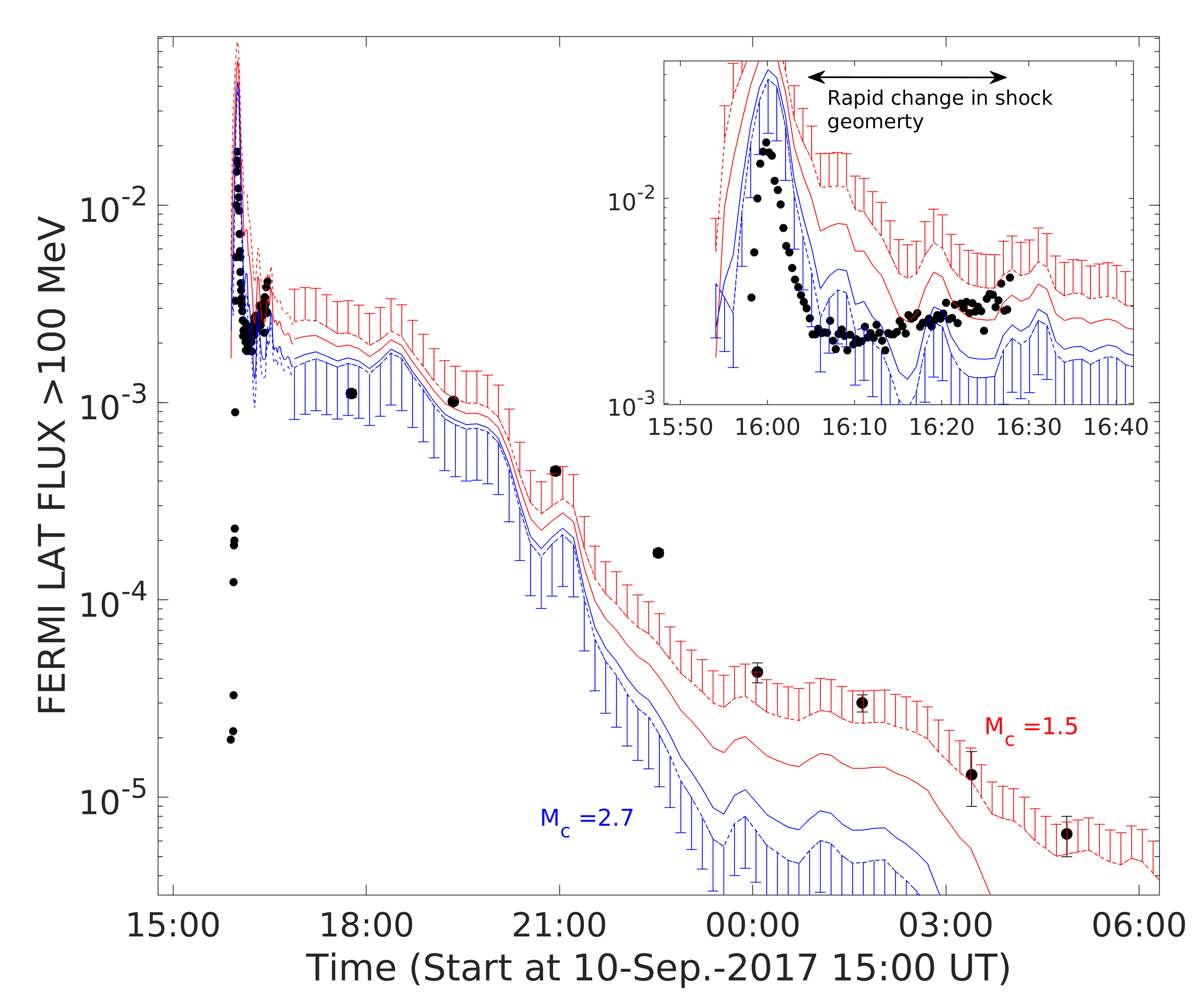}
\caption{Comparison of the evolution of the empirical quantity $\mathrm{F'}$ (colored curves) and the observed $>$100~MeV $\gamma$-ray emission from LAT. The different curves of $\mathrm{F'}$ have been calculated using different values of the critical Mach number, e.g. $\mathrm{M_c: 1.5, 2.0, 2.5, 2.7}$. The inset panel presents a zoomed-in version during the first stage of the late-phase emission from 15:50~UT to 16:40~UT. 
\label{fig:Gflux_FMA} }
\end{figure}

In Figure~\ref{fig:Gflux_FMA} we show the comparison of $\mathrm{F'}$ and the observed $\gamma$-ray flux. For the different curves of $\mathrm{F'}$ shown in this panel we have used four different values of $\mathrm{M_c}$, from 1.5 to 2.7, to perform the calculation. Considering that the shock geometry changes from quasi-perpendicular to quasi-parallel (see Figure~\ref{fig:GRAYS2ndProp}) we expect that the curves calculated for higher $\mathrm{M_c}$ values would better match the $\gamma$-ray fluxes before 16:08 UT and that lower $\mathrm{M_c}$ values would begin to match better after that time. This appears to be happening in the curves plotted in the inset of Figure~\ref{fig:Gflux_FMA} where the $\gamma$-ray fluxes follow $\mathrm{F'}$ for $\mathrm{M_c \sim 2.7}$ until about 16:08 and when they rise above that curve and better match the $\mathrm{F'}$ curve for $\mathrm{M_c \sim 1.5}$ at 16:25 UT.  

The empirical acceleration parameter $\mathrm{F'}$ tracks the $\gamma$-ray fluxes fairly well over the full event, in contrast to the large disparity in values of the shock parameters relative to the $\gamma$-ray fluxes between 16 UT and 18 UT plotted in Figure~\ref{fig:GRAYSPropExtent}. This is likely due to the fact that $\mathrm{F'}$ takes into account both geometric effects of shock strength and the efficiency of particles to reach the solar atmosphere due to magnetic mirroring.  However, there still are intervals where the model and the data diverge. This is due to the basic nature of the $\mathrm{F'}$ factor, which fails to fully account for the physical conditions. For example, we have not taken into account changes in the density of seed particles. We have also not taken into account the complicated transport of particles in the downstream direction from the shock to Sun where the density is higher and significant wave turbulence is present. This is especially important after 16:30 UT when all the magnetic connections to the visible solar disk are downstream of the shock.

Another important simplifying factor is that we assumed that scattering was not significant in determining $\mathrm{F'}$. If there is magnetic turbulence the fraction of particles reaching the solar atmosphere would rise significantly. When the scattering mean free path is small, the particles can continuously scatter and eventually enter the loss cone \citep[e.g.][]{Effenberger2018}. We estimate if this effect is important during our modeling interval by performing a similar calculation to \cite{Jin2018} for the $\mathrm{T_{esc} / \Delta T }$, i.e. the ratio between the particle escape time from the magnetic trap, $\mathrm{T_{esc}}$, to the duration $\mathrm{\Delta T }$ of the emission. This ratio is analogous to $\mathrm{T_{esc} / \Delta T \propto 2 \eta \upsilon_{sh}/c}$ and when $\mathrm{T_{esc} \lesssim \Delta T }$ we expect the number of particles reaching the solar atmosphere within the emission duration $\mathrm{\Delta T }$ to rise significantly. For a shock speed of $\mathrm{\sim3000~km\,s^{-1}}$ we find that this would happen when $\mathrm{\eta\le}$50. From our modeling we have that $\langle\eta\rangle$ is less than 50 during the first stage, before $\sim$16:09~UT. Thus, we would expect that more particles were able to reach the solar atmosphere during the first tens minutes of the event than is inferred from the $\mathrm{F'}$ parameter alone.

\subsection{Can Shock Acceleration Explain the Gamma-Ray Source Localization?}

\begin{figure*}[ht!]
\centering
\includegraphics[scale=0.70]{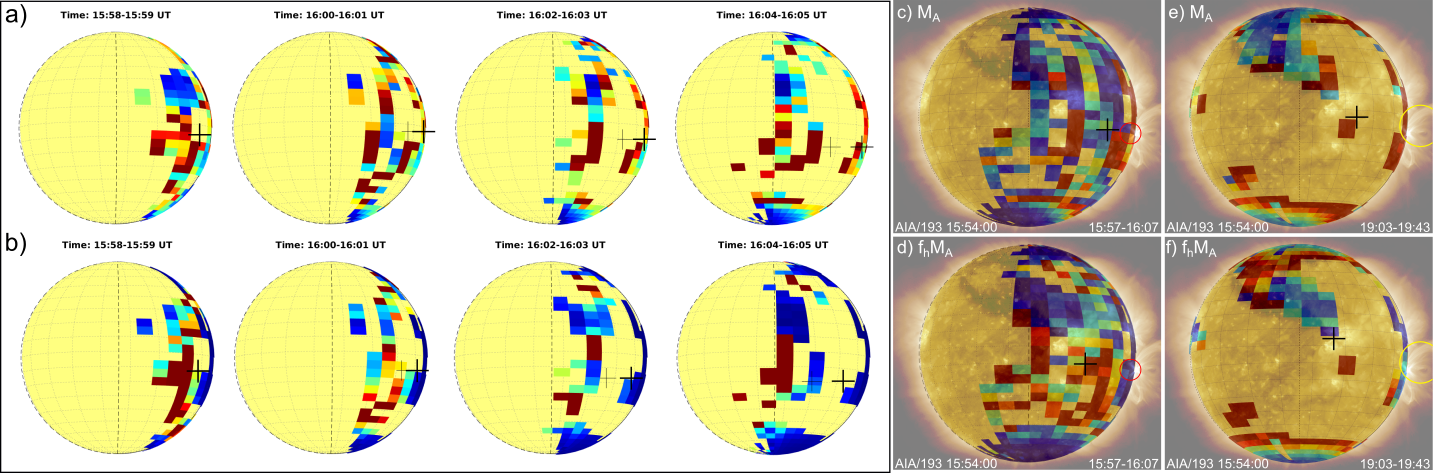}
\caption{ Localization analysis based on the $\mathrm{M_A}$, in the time interval from 15:57 to 16:07~UT. Panel (a) shows the maps of $\mathrm{M_A}$ in two-minute intervals from 15:58 to 16:05 UT, and (b) shows the same maps corrected for magnetic mirroring. Panel (c) shows the map of $\mathrm{M_A}$ integrated over the full selected time interval, and (d) the same map corrected for magnetic mirroring. Panels (e) and (f) show the maps integrated over the full interval from 19:03 to 19:43UT. The black cross in each map is the intensity-weighted average of the parameter distribution. An EUV image at 15:54:00~UT from SDO/AIA observations at 171$\mathrm{\AA}$ is overlaid. The red and yellow circles shows the 95\% confidence $\gamma$-ray location during the selected time intervals \citep{Omodei2018}. 
\label{fig:GLocal} }
\end{figure*}

\cite{Omodei2018} have determined the location of the centroid of the $>$100 MeV $\gamma$-ray emission observed over three different time intervals during the event. They are all consistent with the location of the flare and active region. The most accurate centroid position, determined between 15:57 and 16:07 UT, was 80 degrees $\pm$90 arcsec from the AR. We compare this location with the maps of shock parameters magnetically connected to the visible disk. Such maps for a one-minute interval at 16:00 UT are plotted in panels g and h of Figure~\ref{fig:3DShockProp} for the Alfv\'enic Mach number and density compression ratio. 

In panel (a) of Figure~\ref{fig:GLocal} we plot the maps of Alfv\'enic Mach number in two minute intervals from 15:58 to 16:05 UT. As discussed in Section~\ref{sec:ShockModel}, early in the event a significant fraction of the field lines are closed with footpoints near the active region and in expanding rings as the shock encounters larger magnetic loops. Strong shock regions are connected close and around the AR. The black crosses show the intensity-weighted centroids of the Alfv\'enic Mach number for the full distribution. It is clear that the centroid moves eastward of the active region with time. In panel (c) we show the map of the Alfv\'enic Mach number covering the full 15:57-16:07 UT time interval studied by \cite{Omodei2018}. The LAT 95\% confidence location is shown by the open circle. The intensity-weighted centroid is located close to, but not consistent with, the location of the centroid of the observed $\gamma$-ray emission.

For estimating the impacting particle distribution, we would also need to take into account the amount of magnetic mirroring and turbulence at each of the footpoints. We can correct the maps for magnetic mirroring using the method described in Section~\ref{sec:MagnetRatio}. These corrected maps are shown in panel (b) of Figure~\ref{fig:GLocal}. We see that the maps of Alfv\'enic Mach number with magnetic mirroring produce centroid locations further East of the gamma-ray location (see also panel (d)). The only way for the shock model to produce a particle distribution that is consistent with the LAT centroid location is if the particles only precipitate into the solar atmosphere near the active region. This can occur only in the presence of strong magnetic turbulence associated with the eruptive event. \cite{Omodei2018} also determined the centroid of gamma-ray emission later in the event from 19:03 to 19:39 UT. We plot in panels (e) and (f) of Figure~\ref{fig:GLocal} the maps of Alfv\'enic Mach number (with and without accounting for magnetic mirroring) covering the interval from 19:03 to 19:43 UT. During this time interval the magnetic connections from the shock are on open field lines connected to the AR and to polar coronal holes. Significant magnetic turbulence must prevail in the AR in order for the modeled location to agree with the observations.

\section{Connecting the shock evolution with the time history of the relativistic protons at Earth} \label{sec:GLEShock} 

\begin{figure*}[ht!]
\centering
{\includegraphics[scale=0.70]{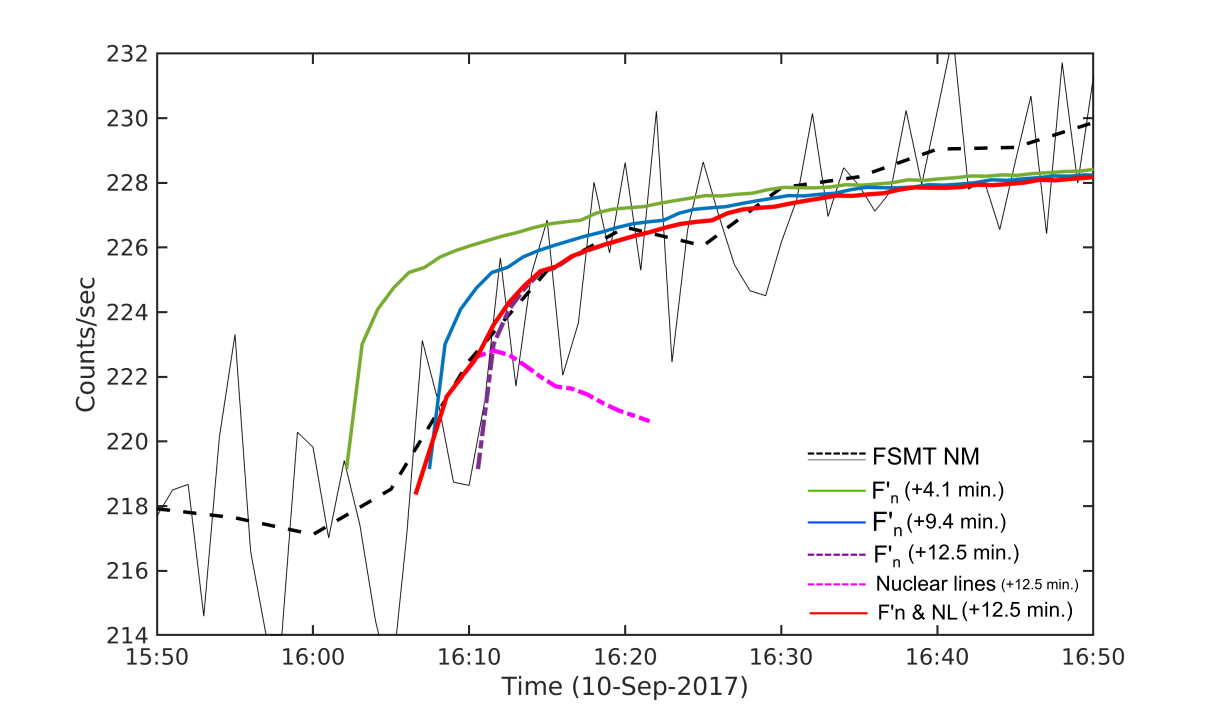}}
\caption{ Comparison between the evolution of the $\mathrm{F'_n}$ and the time history of the GLE72 count rates from FSMT neutron monitor (see text for details). \label{fig:GLE72_multi} }
\end{figure*}

In this section we investigate whether particles accelerated by shocks crossing field lines reaching Earth can explain the time history of relativistic protons observed in GLE72. \cite{Kurt2019} detail observations made by several neutron monitors. They show that the proton angular distribution was strongly anisotropic early in the event and only became isotropic at about 18:00 UT. The Fort Smith (FSMT) neutron monitor located in the Northwest Territories (60 North, 112 West) was relatively closely aligned with solar magnetic field lines at 16:00 UT. Therefore it provides an excellent monitor of early particle acceleration at the Sun. We plot the pressure corrected FSMT 1-min and 5-min count rates from 15:30 to 17:00 UT in Figure~\ref{fig:GLE72_multi}. The rates rise in response to the event at about 16:08:30 UT.

In determining the evolution of the shock wave parameters at regions well-connected to Earth we use the methods presented in \cite{Kouloumvakos2019} and also in our discussions in Sections~\ref{sec:ShockModel} and~\ref{sec:GrayShock}. The shock wave intersected magnetic field lines that were well-connected to Earth at 15:57 UT and it became super-critical around 15:59 UTl\footnote{This assumes that the shock accelerating the particles was quasi-perpendicular requiring a critical Mach number of $\sim$2.7 which was reached at 15:59 UT.}. The mean $\mathrm{M_{A}}$ and $\mathrm{M_{fm}}$ parameters continued to increase, reaching a maxima near 18:00. The mini-neutron monitor DOMC \citep{Usoskin2015} NM station situated at the South Pole recorded the most intense flux with a peak at 18:50~UT  \citep[see][]{Mishev2018}. The density compression ratio saturated within 15~minutes and reached an asymptotic value of four.

Using a similar model to the one used in Section~\ref{sec:MagnetRatio} to predict the time profile of the $\gamma$-ray emission, we examine the shock origin of the particles that access field lines reaching Earth. We consider an $\mathrm{F'_n}$ function that includes only the term, $\epsilon_{sh}$ discussed in Section~\ref{sec:MagnetRatio}, that approximates the effectiveness of shock acceleration. In order to directly compare estimated proton flux with the FSMT NM count rate, we need to shift $\mathrm{F'_n}$ to account for the time it took the protons to reach Earth. A common approach is to assume that the particles propagate along IP magnetic field lines following a Parker spiral whose shape is defined by the $\mathrm{500~km\,s^{-1}}$ solar wind speed at the time of the event\footnote{Measured by the Advanced Composition Explorer (ACE)/Solar Wind Electron, Proton, and Alpha Monitor (SWEPAM)}. For this wind speed, we estimate that the protons traveled 1.1 AU to reach Earth. As neutron monitors typically respond to protons with energies above about 450 MeV \citep{Clem2000}, we assume that the particle velocity was 0.737c. For this velocity and path length the protons would have arrived at Earth with a delay of 4.08 minutes relative to the electromagnetic radiation emitted at the same time. The green curve in Figure~\ref{fig:GLE72_multi} plots our estimate of the proton time history at Earth for this delay. Although the shape of the time history is similar to the neutron monitor count rate, the delay appears to be about 7~minutes too small.

Previous studies have required longer path lengths than provided by a simple Parker spiral during SEP and GLE events \citep[e.g.][]{Masson2012,Vainio2013,Laitinen2019} due to IP disturbances or turbulent magnetic fields \citep[e.g.][]{Laitinen2019}. There were several CME events in the previous days preceding the September 10 flare. Such ICMEs can distort the IP magnetic field. If a magnetic cloud from a previous event was near Earth during the GLE72 the protons could have streamed along helical field lines in the cloud, thus significantly increasing their path lengths. Following arguments in \cite{Rouillard2016}, a toroidal CME flux rope could enclose quasi-circular field lines passing along the center of the toroid yielding lengths as long as 1.57 AU and a delay of 9.39~minutes (blue curve in Figure~\ref{fig:GLE72_multi}). Pitch angle scattering of the protons in the turbulent IP medium can add an additional 0.2~AU onto the pathlength, yielding a length of about 1.8 AU. For this pathlength, protons with velocity of 0.737c would arrive about 12 minutes after the electromagnetic radiation. 

It is clear from the figure that the shock-produced time profiles can fit the FSMT rates for particle particle delays in excess of 10 minutes. We attempted to make a quantitative estimate of how well the shock model fit the data and the best particle delay. In order to account for a flare contribution to the FSMT rate, we assumed that these protons were released into space with the time profile following that of nuclear deexcitation lines plotted in Figure~\ref{fig:FLATLC_b}c. We assumed that both the flare and shock protons were delayed by the same amount and left their relative intensities as a free parameter in our study. We performed what amounted to a chi-squared minimization analysis and found a particle delay of 12.5 $\pm$3 minutes. A significant contribution from flare particles appears to be needed before 16:12 UT. The best-fitting flare and shock components and their total are plotted in Figure~\ref{fig:GLE72_multi}. From about 16:07 to 16:17 UT, we estimate that this impulsive component contributed about 29\% $\pm$8\% of the observed neutron monitor counts.

\section{Discussion and Conclusions} \label{sec:Discussion}

There is growing evidence that LPGRE, which has now been observed for over twenty-five years is in large part produced by protons accelerated by the same CME shocks that are believed responsible for SEPs. This is based on both historical studies and more recent analyses of heliospheric data related to a large sample of events observed by the Fermi/LAT detectors \citep{Plotnikov2017,Gopalswamy2018b,Jin2018,Kahler2018,Share2018,Winter2018}. However, there are also studies that raise concerns about such an origin \citep{Hudson2018,deNolfo2019}. Data from the eruptive limb event on 2017 September 10 \citep{Omodei2018} offer the opportunity to critically assess the validity of the CME/shock origin for LDGRE. This eruptive event also produced a ground level enhancement in solar energetic particles with a distinct early time profile that can also be used to test the shock origin of this emission. 

The 2017 September 10 eruption was the most intense solar event observed by Fermi, and its complex time profile represents a challenge for any model. The $>$100~MeV $\gamma$-ray time history appears to consist of three stages: a weak early impulsive phase that follows the time profile of the impulsive nuclear $\gamma$-ray line emission, and a second phase that consists of two stages: an intense ten minute stage (with a peak much stronger than that of the weak impulsive-phase emission), and an extended stage peaking about two hours later and lasting for close to 14 hours. We use a variety of data products to model the shock wave produced by the expanding CME and we determine the time history of the shock parameters at times when they encounter magnetic field lines returning to the visible disk of the Sun as viewed from Earth.

We show that the two stages of late-phase $\gamma$-ray emission are a reflection of the times when the shock parameters exceed the critical level for particle acceleration and of the coronal conditions that determine the magnitude of the shock strength as it moves through the corona. The calculated time history of the shock parameters mirrors the two stages of LPGRE. We find that the temporal agreement between the observed $\gamma$-ray fluxes and our predicted rates improve significantly when we take into account both the efficiency of particle acceleration with change in shock angle relative to the field lines and the magnetic mirroring that prevents the protons from reaching dense regions in the solar atmosphere where they can interact. We find that the apparent factor of five drop in the $\gamma$-ray flux between the first and second LAT solar exposures during the event primarily arises due to a significant increase in the magnetic mirroring ratio for protons returning to the Sun between the two exposures. This good agreement between the time profile produced by our relatively simple shock model and that observed in $>$100~MeV $\gamma$-rays is compelling evidence for a shock origin for the LPGRE.

Our modeling allows us to determine the spatial distribution of shock-accelerated particles returning to the Sun. During the first ten minutes of the event when the $\gamma$-ray emission peaks and subsides rapidly, the shock parameter distribution over the visible disk is dominated by a ring that increased in size with time and a more concentrated structure near the active region. This reflects the fact that particle acceleration was taking place along both the nose and flank of the shock onto open and closed field lines emanating out from the active region. Strong shock regions are mainly connected to regions near the active region. When magnetic mirroring is taken into account, the overall intensity of particles reaching the solar atmosphere drops significantly but the spatial distribution does not change radically. During this same ten minute interval Fermi/LAT found that the centroid of the $\gamma$-ray emission was consistent with the location of the active region on the limb \cite{Omodei2018}. Such a location on the limb is inconsistent with our calculated distribution, unless magnetic turbulence is taken into account. We would expect there to be high turbulence near the active region \citep[e.g.][]{Ryan1991,Effenberger2018} in comparison with the remote footpoints of the closed loops. Such turbulence causes pitch angle scattering, resulting in particle precipitation into the loss cone and a significantly brighter source near the active region. By 19~UT most of the particle acceleration takes place near the nose of the shock in a quasi-parallel geometry with open field lines emanating from the AR and from coronal holes in the polar regions of the Sun. At that time, \cite{Omodei2018} found a gamma-ray source location once again consistent with the AR on the limb. Thus, in order for the shock model to produce such a location, there still needs to be significant magnetic turbulence near the AR.

We have also modeled shock parameters on field lines reaching the Earth during the first hour of the GLE observed by neutron monitors. This produced a time history of particle emission that we compared with observations of the highly anisotropic radiation made by the magnetically well-connected Fort Smith neutron monitor. In making this comparison we also included an impulsive flare component of the GLE with a release time profile given by nuclear line emission. We fit the data with the particle delay (for 450 MeV protons) and the relative intensity of the shock and flare components as free parameters. We obtained a good fit to the data with a particle delay of 12.5 $\pm$3 min. Such a delay requires a path length of about 1.85 AU indicating that there was a high degree of IP disturbances and magnetic turbulence at the time. Such disturbances are expected \citep[e.g.][]{Masson2012,Vainio2013,Laitinen2019} due to the high level of flaring in the days prior to the event. Our fits also require a significant contribution of impulsive flare particles beginning at 16:07 UT and for about four minutes before the shock-accelerated particles. From about 16:07 to 16:17 UT we estimate that this impulsive component contributed about 29\% of the observed neutron monitor counts.

In summary, we have shown that shock acceleration of particles onto magnetic field lines returning to the Sun and reaching Earth reproduces both the complex hours-long time history of Late Phase Gamma Ray Emission and most of the early time history of GLE particles recorded by a magnetically well-connected neutron monitor. We also find evidence for a weak contribution of flare-produced $>$100 MeV $\gamma$-rays beginning prior to the onset of the LPGRE. In order for the maps of shock parameters magnetically connected to the visible disk to agree with the centroid locations of $\gamma$-ray emission observed by {\it Fermi} requires a significantly larger degree of magnetic turbulence near the flare active region than at remote footpoints.

Based on a comprehensive comparison of the numbers of $>$500 MeV protons in SEP events observed by {\it PAMELA} and at the Sun, inferred from LPGRE events, \citet{deNolfo2019} recently questioned the CME-driven shock scenario as the source of the $\gamma$-ray emission. They found a high ratio of relativistic proton number at the Sun relative to that in IP space. In lieu of a shock origin, \citet{deNolfo2019} suggest that the LPGRE originates in long turbulent magnetic loops where particles are continuously accelerated by a second-order Fermi mechanism \citep[][]{Ryan1986,Ryan1991}. It is questionable whether such a model can explain the complex temporal evolution of the 2017 September 10 LPGRE event as naturally as we have shown for the CME-shock model.

Some GLE events show a double-peaked time structure \citep{McCracken2012}, with the first peak highly-anisotropic and impulsive (``prompt component'') followed by a second peak less anisotropic and gradual (``delayed component''), such as the GLE event of 2005 January 20. The first ``prompt component'' is commonly related to particle acceleration in the flaring active region in the low corona \citep[e.g.][]{McCracken2008,Masson2009} or in the turbulent/reconnecting current sheet \citep{Guo2018,Kurt2019}, while the origin of the ``delayed component'' is attributed to magnetic reconnection and possibly turbulence in large-scale coronal loops during the development of the CME \citep{Klein2014}. In a temporal study of 12 recent GLE events \citet{Aschwanden2012} concluded that 50\% of the events were accelerated during the impulsive flare phase and further that the prompt component ``is consistent with a flare origin in the lower corona."  In contrast, our study of the 2017 September 10 GLE shows that shock acceleration can be the dominant mechanism even during the prompt phase of the GLE and that, although a flare component is required at the onset of the event, it only constitutes $\sim$29\% of the total emission in the first 10 minutes. Flare accelerated particles could serve as a seed population that is further accelerated by the shock wave \citep[e.g.][]{Petrosian2016,Share2018}.

Our study, revealing a common shock acceleration origin for the high-energy protons responsible for the late-phase $>$100 MeV $\gamma$-ray emission and the GLE72 event needs to be expanded to other LPGRE and high-energy SEP events. In particular it is important to study the LPGRE and GLE event on 2012 May 17 in order to explain the relatively low $\gamma$-ray flux in that event and the 2012 March 13 SEP event with no accompanying LPGRE (see the catalog in \cite{Share2018}). Detailed study of the shock and magnetic field configuration during the 2011 March 7 LPGRE event is required to explain the lack of detectable SEP protons $>$100 MeV. This was cited by \citet{deNolfo2019} as evidence against a shock origin. There are several LPGRE events with impulsive emission distinct from the flare that also need to be studied for association with shock acceleration. These include the events on 2011 August 9, September 6 and 24, 2012 January 23, March 4 and 7, June 3, November 27, 2013 April 11, May 14, October 11 and 2015 June 15.   

Note Added in Proof: We learned about a detailed IP transport analysis of the 2017 September 10 event by \cite{Kocharov2020} published after submission of our manuscript. This analysis is based on the assumption that the GLE protons are released from two distinct compact sources located 0.1 to 1 solar radii above the photosphere. Unfortunately, \cite{Kocharov2020} interpreted the Fermi-LAT localization uncertainty of the centroid of the gamma-ray emission as the spatial extent of the source. In fact, the point spread function of the LAT is about four solar radii at 1 GeV.  Based on this assumption, they compared their estimated GLE particle release times to the observed $>$100 MeV gamma-ray time history. In contrast, we show in this paper that the complicated gamma-ray time profile is naturally explained by the CME-shock acceleration of particles onto field lines returning to the Sun and that the same process can explain the early time profile of the GLE.

\acknowledgments

The IRAP team acknowledges support from the French space agency (Centre National des Etudes Spatiales; CNES; \url{https://cnes.fr/fr}) that funds activity in plasma physics data center (Centre de Données de la Physique des Plasmas; CDPP; \url{http://cdpp.eu/}) and the Multi Experiment Data \& Operation Center (MEDOC; \url{https://idoc.ias.u-psud.fr/MEDOC}),  and the space weather team in Toulouse (Solar-Terrestrial Observations and Modelling Service; STORMS; \url{https://stormsweb.irap.omp.eu/}). This includes funding for the data mining tools AMDA (\url{http://amda.cdpp.eu/}), CLWEB (\url{clweb.cesr.fr/}), and the propagation tool (\url{http://propagationtool.cdpp.eu}). A.K. and Y.W. acknowledge financial support from the ANR project SLOW{\_}\,SOURCE, $ANR-17-CE31-0006-01$, COROSHOCK, and FP7 HELCATS project \url{https://www.helcats-fp7.eu/} under the FP7 EU contract number 606692. G.H.S. acknowledges partial support from NRL under Praxis contract N00173-14-C-2027/21403-PXI-004 and the CSRS, and thanks Kim Tolbert of NASA GSFC for providing special Fermi data products. R.M. was supported by the Chief of Naval Research (CNR). This work would not have been possible without the Fermi-LAT and GBM data products and calibrations that are publicly accessible from NASA. We thank the STEREO: SECCHI, S/WAVES; SOHO; LASCO; and SDO: AIA teams and Predictive Science Inc. for publicly providing the data used in this study. The STEREO SECCHI data are produced by a consortium of RAL (UK), NRL (USA), LMSAL (USA), GSFC (USA), MPS (Germany), CSL (Belgium), IOTA (France), and IAS (France). SOHO is a mission of international cooperation between ESA and NASA. The SDO/AIA data are provided by the Joint Science Operations Center (JSOC) Science Data Processing (SDP). We acknowledge NMDB database (http://www.nmdb.eu), founded under the European Union's FP7 programme (contract No. 213007) and Dr. Roger Pyle for providing the Fort Smith neutron monitor data. The present work has been also benefited from discussions held at the International Space Science Institute (ISSI, Bern, Switzerland) within the framework of the international team ``High EneRgy sOlar partICle events analysis (HEROIC)" led by Dr. A. Papaioannou.



\end{document}